\documentclass[10pt,twocolumn,letterpaper]{article}

\usepackage{iccv}
\usepackage{times}
\usepackage{epsfig}
\usepackage{graphicx}
\usepackage{amsmath}
\usepackage{amssymb}
\usepackage{array}
\usepackage{pifont}
\usepackage{capt-of,etoolbox}
\usepackage{multirow}
\usepackage{cite}
\usepackage{hhline}
\usepackage{authblk}
\usepackage{bbding}
\usepackage{algorithm}
\usepackage{algorithmic}
\usepackage{xcolor,colortbl}
\usepackage[pagebackref=true,breaklinks=true,letterpaper=true,colorlinks,bookmarks=false]{hyperref}

\usepackage[breaklinks=true,bookmarks=false]{hyperref}

\iccvfinalcopy 

\def\iccvPaperID{6339} 

\ificcvfinal\fi

\begin{document}

\title{Voice2Mesh: Cross-Modal 3D Face Model Generation from Voices}

\vspace{-15pt}
\author{Cho-Ying Wu \: Ke Xu \: Chin-Cheng Hsu \: Ulrich Neumann\\
University of Southern California \\
{\tt\small \{choyingw, kxu47918, chincheh, uneumann\}@usc.edu}
\vspace{-15pt}
}

\maketitle
\ificcvfinal\thispagestyle{empty}\fi

\begin{abstract}
   This work focuses on the analysis that whether 3D face models can be learned from only the speech inputs of speakers. Previous works for cross-modal face synthesis study image generation from voices. However, image synthesis includes variations such as hairstyles, backgrounds, and facial textures, that are arguably irrelevant to voice or without direct studies to show correlations. We instead investigate the ability to reconstruct 3D faces to concentrate on only geometry, which is more physiologically grounded. We propose both the supervised learning and unsupervised learning frameworks. Especially we demonstrate how unsupervised learning is possible in the absence of a direct voice-to-3D-face dataset under limited availability of 3D face scans when the model is equipped with knowledge distillation. To evaluate the performance, we also propose several metrics to measure the geometric fitness of two 3D faces based on points, lines, and regions. We find that 3D face shapes can be reconstructed from voices. Experimental results suggest that 3D faces can be reconstructed from voices, and our method can improve the performance over the baseline. The best performance gains (15\% - 20\%) on ear-to-ear distance ratio metric (ER) coincides with the intuition that one can roughly envision whether a speaker's face is overall wider or thinner only from a person's voice. See our \href{https://github.com/choyingw/Voice2Mesh}{project page} for codes and data.
\end{abstract}

\begin{figure}[bt!]
\begin{center}
\includegraphics[width=1.0\linewidth]{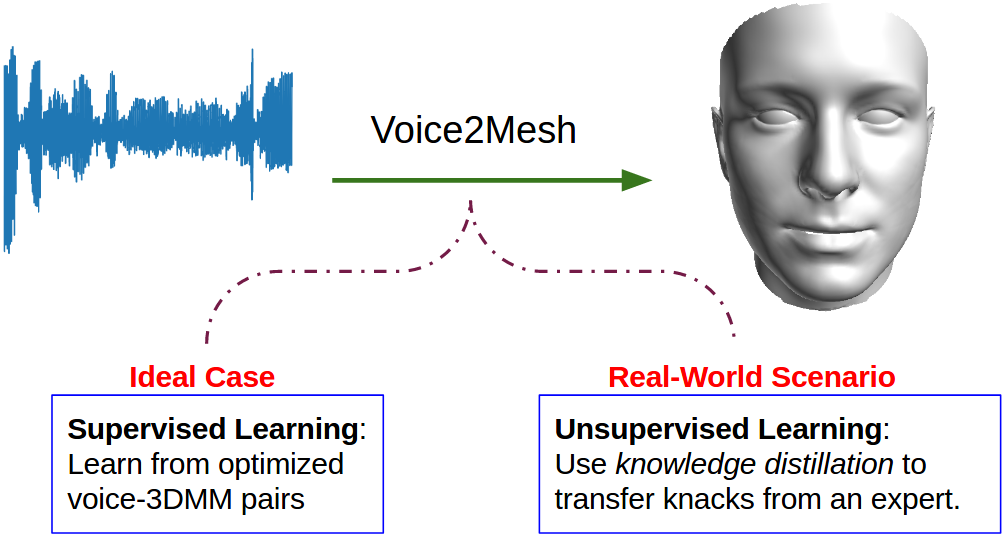}
\end{center}
  \vspace{-6pt}
  \caption{\textbf{Voice2Mesh for both supervised and unsupervised settings.} Our Voice2Mesh recovers a 3D face mesh from a speech input. We present an ideal case and a real-world scenario to deal with the absence of large-scale paired voice and 3D face data.}
\label{purpose}
\vspace{-7pt}
\end{figure}

\section{Introduction}
\label{sec:intro}

Voice is a unique characteristic of humans, and many physiological attributes are embedded in voices; a person's voice is statistically related to their facial structures. 
In physiology, articulatory structures, such as vocal folds, muscles, and skeletons, are densely connected to head and may correlated to the face shapes\cite{harrington2010acoustic}. 
Experiments in cognitive science pointed out that audio cues are correlated with visual cues in human perception-- especially in recognizing a person's identity \cite{belin2004thinking}. 
More recent neuroscience research further shows that two parallel processing of low-level auditory and visual cues are  integrated in the cortex, where voice structural analysis affects facial structural analysis for the perception purpose \cite{young2020face}. 


Traditionally, research in the voice domain focuses on utilizing voice inputs for predicting more conspicuous attributes which include
speaker identity \cite{bull1983voice, maguinness2018understanding, ravanelli2018speaker}, 
age \cite{ptacek1966age, singh2016relationship, grzybowska2016speaker}, 
gender \cite{li2019improving}, 
and emotion \cite{wang2017learning, zhang2019attention}. 
A novel direction of recent development 
goes beyond predicting these attributes and try to reconstruct \textit{2D face images} from voices \cite{NEURIPS2019_eb9fc349, oh2019speech2face, choi2020inference}. 
Their research is built on an observation that one can approximately envision how an unknown speaker looks when listening to voice of the speaker.
Attempts towards validating this assumptive observation include the work by Wen \textit{et al.} \cite{NEURIPS2019_eb9fc349} and Choi \textit{et al.} \cite{choi2020inference} with generative adversarial networks (GANs) and that by Oh \textit{et al.} \cite{oh2019speech2face} with an encoder-decoder network.
These works generate face images from the voices of a speaker.


However, their focus of reconstructing 2D face image from voices inherently suffers from much ambiguity, involving predicting background, hairstyles, beards, etc, that voices cannot hint. These factors are apparently those that one can choose without changing voices.  Likewise, the grounds from physiology only suggest that articulatory and skeletal structures are correlated to voices. Moreover, it is difficult to directly calculate the numerical synthesis error with respect to a reference due to the intrinsic variations. Similar concerns arise regarding the correlation between voice and facial textures or ethnicity, as shown in the t-SNE embeddings graph in \cite{oh2019speech2face}, which shows mixed features for ethnicity.


Instead of reconstructing face images, we propose a new task of \textit{3D face modeling from voices} to validate the correlation between voices and geometry based on the neuroscience support.
Mesh representations of a face are arguably less ambiguous than image representations because the former do not include the noisy variations unrelated to the speaker identity and voice.
The advantage of studying 3D face reconstruction is that it allows us to focus on geometric differences, disregarding stylistic variations such as hairstyles, background, and facial textures.
Moreover, mesh representations enable easier calculation of difference between a reconstruction and a groundtruth, unlike the case in image representations where two images can have different backgrounds and hairstyles.     

From the perspective of 3D face modeling, much research attention has been paid on reconstruction from monocular face images \cite{shang2020self,guo2020towards,zhu2016face} or video sequence \cite{garrido2016reconstruction, kim2018deep} to attain 3D face animation or talking face synthesis. In contrast, we are the first to investigate the cross-modal learning for generating a 3D face model from only speech input, and focus on how much accuracy could be attained for 3D face modeling from voice input. 

However, acquisition of large-scale 3D face scans with paired voices is expensive and subject to privacy. To deal with the issue, we adopt an optimization-based approach \cite{Zhu_2015_CVPR} that the widely-used 3D face dataset 300W-LP-3D \cite{zhu2016face} exploits to perform 3D Morphable Models (3DMM) fitting. We create a paired dataset, Voxceleb-3D, for voices and 3D faces, from the voice dataset (Voxceleb \cite{nagrani2017voxceleb}) and image dataset (VGGFace \cite{BMVC2015_41}) for celebrities.

We propose our main framework \textbf{Voice2Mesh} to investigate the ability to reconstruct 3D face models from voices on the following two scenarios. In our \textit{supervised learning} setting, we base on Voxceleb-3D and train Voice2Mesh directly using the paired voice and 3DMM parameter data.

For a more realistic scenario where a real voice-to-3D-face dataset is unavailable, we investigate the problem under \textit{unsupervised learning} setting. In this case, we transfer the knacks using \textit{knowledge distillation} (KD) \cite{hinton2015distilling} from a state-of-the-art 3D face modeling method, 3DDFA-V2 \cite{guo2020towards}, as the pretrained expert into our student network and jointly train the pipeline of a voice-to-image and an image-to-3D module.
In addition, we also extensively compare the performance of existing KD methods. Illustration in Fig. \ref{purpose} shows the intuition of our Voice2Mesh.


For the evaluations, we design different metrics to measure the geometric fitness based on points, lines, and regions for both the supervised and the unsupervised scenarios. Our purpose is not to synthesize high-quality 3D face mesh from voices that is comparable to synthesis from visual modality such as images or videos. The purpose of the evaluation is to show to what extent 3D facial geometry could be predicted from voices and to analyze how voice information helps predict a 3D face model using KD with a pretrained image-to-3D-face network.

\begin{enumerate}
  \item We are the first to study the cross-modal 3D face modeling from voice input and to analyze to what extent 3D facial geometry information is embedded in voices.
  \item We study both supervised and unsupervised approaches with knowledge distillation for learning 3D facial geometry from voices, considering the scarcity of 3D face scan datasets. 
  \item We perform extensive evaluations for measuring the reconstructed 3D face models from voices to validate the correlations of voice and human face shape.
\end{enumerate}

\section{Related Work}


\subsection{Recognizing Personal Traits with Voice}

Human voice is embedded with a wide range of personal information and has long been exploited for recognizing personal traits, 
such as speaker identity \cite{bull1983voice, maguinness2018understanding, ravanelli2018speaker}, age \cite{ptacek1966age, singh2016relationship, grzybowska2016speaker}, gender \cite{li2019improving}, and emotion status \cite{wang2017learning, zhang2019attention}.
Voices can also be used to monitor health condition \cite{ali2017automatic} or applied to other medical applications \cite{han2021exploring}. 
Most existing works focus on predicting personal traits that are more intuitively related to voice.

\subsection{Face Synthesis} 

Face-related synthesis has been under much research in the past years.
Generating 2D face image using GANs \cite{goodfellow2014generative, abdal2019image2stylegan, karras2019style, choi2018stargan} has been a popular task and recent progresses include more realistic synthesis with diverse  styles. The task of face reenactment \cite{garrido2014automatic, nirkin2019fsgan, thies2016face2face} focuses on transferring facial features from a source to a target. 
Some works focus on 3D domain: synthesizing 3D face models from monocular images \cite{zhu2016face, guo2020towards, tran2018nonlinear}, or synthesizing motions of 3D faces from videos \cite{kim2018deep, garrido2016reconstruction} using 3DMM \cite{egger20203d}.


\subsection{Audio-Visual Learning}

\textbf{Cross-modal Face Matching}. Cross-modal face matching tries to match voice-face pairs with one modality as an input and the other as the fetched result. This task is inherently a \textit{selection} problem that choose the best fit of voice-face pairs from the dataset. \cite{nagrani2018seeing,kim2018learning} 

\textbf{Talking face synthesis}. Talking face generation targets at synthesizing coherent and natural lip movements. Some works synthesize the faces by audio inputs and a template image \cite{jamaludin2019you, zhou2019talking} or a template 3D face model \cite{cudeiro2019capture}. Other works use an audio plus source video as inputs to drive the synthesis for the target identity \cite{chen2018lip, wiles2018x2face}. 
Different from our attempt of voice-to-3D-face prediction, 
these works use a combination of audio, 2D, 3D, and videos for predicting coherent and natural lip movements.

\textbf{Voice to Face}. Voice to face is the closest task to our proposed cross-modal 3D face modeling. This task is introduced recently to synthesize face images from only voice inputs. Wen \textit{et al.} \cite{NEURIPS2019_eb9fc349} and Choi \textit{et al.} \cite{choi2020inference} adopted GANs to generate face images from audio clips. Oh \textit{et al.} \cite{oh2019speech2face} used an encoder-decoder structure to reconstruct face images. However, the disadvantages are that 2D representations contain many variations, such as hairstyles, beard, background, and facial textures that are irrelevant to facial geometry, or the correlations lack physiological support. Besides, face reconstruction error can be ambiguous because two images of the same person can contain different hairstyles and backgrounds.

Our cross-modal 3D face modeling circumvents the issues raised by 2D face representations. First, our 3D face model does not contain hairstyles, backgrounds, or texture variations. Geometric representation of mesh enables us to study the correlation between voices and 3D face shapes. Next, we are able to directly compute the 3D face modeling error using the geometric representation of mesh.

\subsection{Knowledge Distillation for Applications}

Knowledge distillation is introduced to distill neural network information from a teacher or expert model to a student model \cite{hinton2015distilling}. More advanced KD techniques are introduced for better information transfer \cite{tian2019crd, tung2019similarity, peng2019correlation, ahn2019variational,passalis2018learning, huang2017like, zagoruyko2016paying, romero2014fitnets, passalis2020probabilistic}. This technique originally targets at model compression \cite{romero2014fitnets, hinton2015distilling}. Some of the recent works exploit KD for different applications. Gupta \textit{et al.} \cite{gupta2016cross} use KD to transfer the supervision to perform semantic segmentation on depth maps. Afouras \cite{afouras2020asr} adopt KD to distill information from a teacher model of auditory speech recognition to a student model of visual-speech recognition for lip reading. Multi-Modal or cross-modal learning usually encounters missing data issue for some modalities, and thus the supervised learning is not feasible. Instead, some works adopt KD to distill information from experts to facilitate unsupervised learning \cite{weinzaepfel2020dope, kothandaraman2020unsupervised}. In our unsupervised setting, we also utilize KD to fulfill our training pipeline for voice to 3D face modeling.


\section{Methods}
\label{sec:methods}
Our goal is to analyze how human voices are correlated to 3D facial geometry. Therefore we propose Voice2Mesh, which learns 3DMM parameters from speech inputs. We first review 3DMM and then introduce our Voice2Mesh under two learning schema, supervised and unsupervised fashion. The supervised learning is an ideal case when an paired voice and 3D face dataset is available. The unsupervised learning is to deal with the absence of such paired dataset. The logics behind the division is illustrated in Fig.\ref{purpose}.

\subsection{3D Morphable Models (3DMM)}

3DMM \cite{egger20203d} is a popular framework for 3D face modeling using principal component analysis (PCA). By estimating PCA weights for variation subspaces for \textit{facial shape} and \textit{expression}, and combining the weights with basis matrices, 3D face models can be constructed. 

We formalize 3DMM face modeling as follows. Suppose we have an average face $\bar{A} \in \mathbb{R}^{3N}$ with $N$ three-dimensional vertices, a basis matrix of the shape variational subspace $V_{s} \in \mathbb{R}^{3N\times P_s}$, the corresponding coefficients $\alpha_s \in \mathbb{R}^{P_s}$, a basis matrix of the expression subspace $V_{e} \in \mathbb{R}^{3N\times P_e}$, and the corresponding coefficients $\alpha_{e}\in \mathbb{R}^{P_e}$.
A 3D face can be constructed by the following relations:
\begin{equation}
     A = \bar{A} + V_{s}\alpha_{s} + V_{e}\alpha_{e},
\label{3DMM_basics}
\end{equation}
where $A \in \mathbb{R}^{3N}$ is a vector. We can reshape $A$ into a matrix representation $A_r \in \mathbb{R}^{3\times N}$ showing $N$ points 3D vertices.

We then adopt BFM \cite{paysan20093d}, a particular form of 3DMM that is widely used. It specifies $N=53490$ vertices to represent a face mesh. For the subspaces, we follow the state-of-the-art work 3DDFA-V2 \cite{guo2020towards} and specify $P_s=40$ and $P_e=10$.
Note that 3DDFA-V2 also predicts 12-dimension pose parameters (a $3\times 3$ matrix for rotation and a 3D translation vector). Then the 3D face model can be transformed by $A_p = RA_r+t$ to the predicted pose. However, in our Voice2Mesh, we focus on the deformation of the 3D face model from voice input and thus do not predict pose.


\subsection{Voice2Mesh - Supervised Learning}
\label{sec:methods/supervised}
 
\begin{figure}[bt!]
\begin{center}
\includegraphics[width=1.0\linewidth]{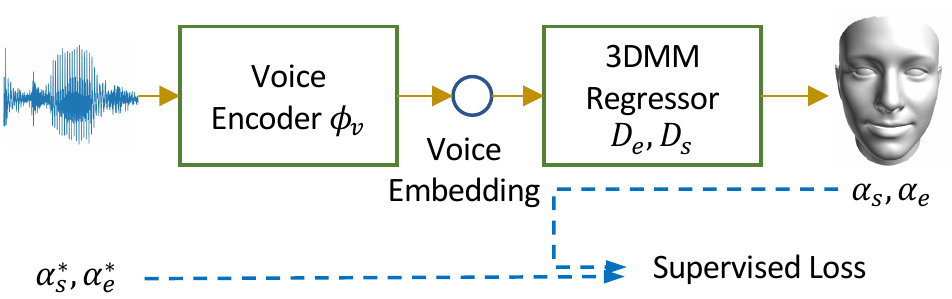}
\end{center}
  \vspace{-7pt}
  \caption{\textbf{Voice2Mesh of supervised learning.} Given the input speech $S$, voice embedding extractor and $\phi_v$ extract voice features and decoders $D_s$ and $D_e$ regress 3DMM parameters $\alpha_s$ and $\alpha_e$ for shape and expression.}
\label{supervised_fig}
\vspace{-10pt}
\end{figure} 
 
For the supervised learning, given paired voice and 3DMM parameter data, we build an encoder-decoder network to first extract voice embeddings $v\in \mathbb{R}^{64}$ from mel-spectrogram \cite{grochenig2001foundations}, which is a commonly used time-frequency representation for speech, of the input speech sequence $S$. Following \cite{NEURIPS2019_eb9fc349}, the voice embedding extractor $\phi_v$ is pretrained on a large-scale speaker recognition dataset. Then, we train separate decoders $D_s$ and $D_e$ to regress 3DMM parameters for $\alpha_s$ and $\alpha_e$. The network structure is illustrated in Fig. \ref{supervised_fig}.

We use groundtruth 3DMM parameters to supervise the training using L2 loss.
\begin{equation}
      \mathcal{L}_{super} = \|\alpha_{s}-\alpha^*_{s}\|^2 + \|\alpha_{e}-\alpha^*_{e}\|^2,
\label{supervised_loss}
\end{equation}
where $\alpha^*_s$ and $\alpha^*_e$ are groundtruth parameters for shape and expression.

The challenge for the supervised learning is for obtaining $\alpha^*_s$ and $\alpha^*_e$. Voice dataset as Voxceleb \cite{nagrani2017voxceleb} contains only speech for celebrities. Face dataset as VGGFace \cite{BMVC2015_41} only contains publicly scraped celebrity face images. We follow \cite{NEURIPS2019_eb9fc349} to fetch the intersection of voice and image data from Voxceleb and VGGFace. Next, we adopt an optimization-based approach \cite{zhu2016face} that the most widely used 3D face dataset, 300W-LP-3D, uses. We use an off-the-shelf 3D landmark detector \cite{bulat2017far} to extract facial landmarks from collected face images and then optimize 3DMM parameters to fit in the extracted landmarks. In this way, we obtain paired voice and 3D face data to fulfill the supervised learning.

\subsection{Voice2Mesh - Unsupervised Learning}
\label{sec:methods/unsupervised}

The purpose to study the unsupervised learning is that obtaining 3D face scans is very expensive and also limited by the privacy issue. 
The workaround with optimization-based methods to fit facial landmarks to 3DMM parameters is time-consuming. 
As a result, the value of unsupervised version of Voice2Mesh lies in studying how 3D faces could constructed from voice inputs without groundtruth 3DMM parameters or 3D face models, which is closer to real-world scenarios.

Our unsupervised Voice2Mesh includes two stages, face image synthesis from voices using GANs and 3D face modeling from generated face images. After image synthesis from the generator of GAN, we build an 3D face reconstruction network to recover the 3D facial structure from monocular synthesized face images. The motivation is that 3D face reconstruction network from monocular images disentangles the geometric representation of faces from other variations such as hairstyles, facial textures, and backgrounds in images. Therefore it is desirable to extract out 3D facial geometry from generated images and exclude the irrelevant variations with voices.

\textbf{Face image synthesis from voices with GANs.}
Previous research develops a GAN-based voice-to-image face synthesis framework \cite{NEURIPS2019_eb9fc349}. A voice encoder $\phi_v$ extracts voice embeddings from input voice. Then a generator $\phi_g$ synthesizes face image from the voice embeddings, and a discriminator $\phi_d$ decides whether the synthesis is indistinguishable from a real face image. The generator and the discriminator forms a GAN that can generate higher quality of face images. Lastly, a face classifier $\phi_c$ learns to predict the identity of an incoming face, making sure that the generator produces face images that are truly close to the face image of the person in interest. We slightly abuse the notation $\phi_v$ and other later introduced ones for 3D face modeling in both Sec.\ref{sec:methods/supervised} and \ref{sec:methods/unsupervised} due to the same functionalities.


Suppose given a speech input $S$ and the corresponding speaker ID $id$ and real face images for the speaker $I_r$, the image synthesized from the generator is $I_f = \phi_g(\phi_v(S))$. The loss formulation is divided into two parts: real and fake images. For real images, the discriminator learns to assign them to "real" ($r$) and the classifier learns to assign them to $id$. The loss for real images is $\mathcal{L}_r = \mathcal{L}_d(\phi_d (I_r), r)+\mathcal{L}_c(\phi_c (I_r), id)$ showing the discriminator and classifier losses respectively. For fake images, after producing $I_f$ from $\phi_g$, the discriminator learns to assign them to "fake" ($\bar{r}$) and the classifier also learns to assign them to $id$. The loss counterpart for fake image is $\mathcal{L}_f = \mathcal{L}_d(\phi_d (I_f), \bar{r})+\mathcal{L}_c(\phi_c (I_f), id)$.

\textbf{3D face modeling from synthesized images.} After obtaining synthesized images from the generator, we build a parameter regression network to learn 3DMM parameters from fake images. The parameter regression network consists of an encoder $\phi_{I}$ and an decoder $D_s, D_e$ to regress shape and expression parameters $\alpha_s = D_s(\phi_{I}(I_f))$ and $\alpha_e = D_e(\phi_{I}(I_f))$. 3D face mesh models are then constructed by Eq.(\ref{3DMM_basics}).

\textbf{Knowledge distillation for unsupervised learning}
\label{sec:unsuper}

\begin{figure}[bt!]
\begin{center}
\includegraphics[width=1.0\linewidth]{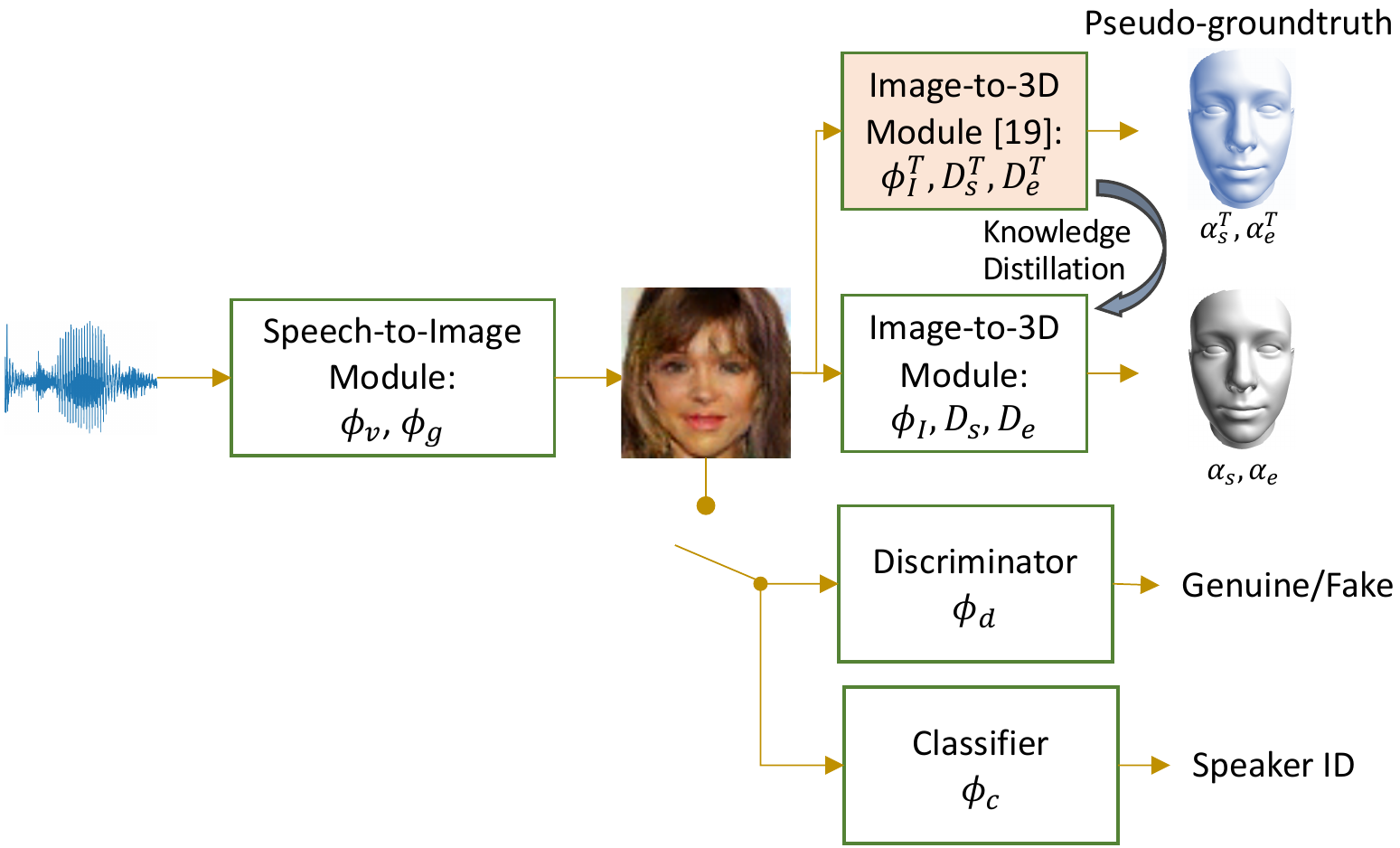}
\end{center}
  \caption{\textbf{Voice2Mesh of unsupervised learning.} The unsupervised learning framework contains a voice encoder $\phi_v$, generator $\phi_g$, discriminator $\phi_d$, classifier $\phi_c$, for face image synthesis using GAN. KD is utilized to attain unsupervised learning where the information distilled from the teacher or expert network is exploited to train the student network for 3D face modeling from synthesized face images as the latent representation. Beside using pseudo-groundtruth $\alpha_s^T$ and $\alpha_e^T$ to train the student, we also distill knowledge of intermediate layers from the teacher to student. The pink background refers to fixed pretrained module.}
\label{kd_pipeline}
\end{figure} 

To deal with the availability issue for 3D face scans, we use KD to distill knowledge from a pretrained teacher network for image to 3D face reconstruction. In this way, the \textit{paired voice and 3D face scan data are not required}. The distilled information from the teacher network helps train the student network to predict 3DMM parameters for 3D face modeling. The framework is illustrated in Fig. \ref{kd_pipeline}.

A teacher network for 3D face modeling, consisting of encoder $\phi_{I}^T$ and decoder $D_s^T, D_e^T$, helps reconstruct 3D face models for the intermediate face image and produce pseudo-groundturh of 3DMM parameters $\phi_s^T$ and $\phi_e^T$. The pseudo-groundtruth is used t supervise the training of student. The loss is formulated as follows. 
\begin{equation}
     \mathcal{L}_{p-gt}=\|\alpha_{s}^T-\alpha_{s}\|^2 + \|\alpha^T_{e}-\alpha_{e}\|^2. 
\label{loss_pgt}
\end{equation}

In addition to pseudo-groundtruth, we also constrain on the probability distribution of intermediate layer outputs and minimize the distribution divergence. We measure the probability distribution from the feature space of the extracted image embedding $z^T \in \mathbb{R}^{B \times \nu}$ and $z \in \mathbb{R}^{B \times \nu}$ of the teacher and the student network. The image embeddings keep the batch dimension $B$ and collapse the other dimensions to $\nu$.

We further calculate the conditional probability used in \cite{passalis2020probabilistic}, which represents the probabilities for selecting neighbors of each feature point. 
\begin{equation}
     z_{i|j}=\frac{K(z_i,z_j)}{\sum_{k,k\ne j}K(z_k,z_j)},
     z^T_{i|j}=\frac{K(z_i^T,z_j^T)}{\sum_{k,k\ne j}K(z^T_k,z^T_j)},
\label{loss_pgt}
\end{equation}
where $K(\cdot, \cdot)$ is a symmetric kernel function. We use scaled and shifted cosine similarity to make the function output lies in [0,1]. 

\begin{equation}
     K_{cosine}(z_i,z_j)=\frac{1}{2} \left(\frac{z_i^T z_j}{\|z_i^T\|_2 \|z_j\|_2}+1 \right).
\label{loss_pgt}
\end{equation}

Note that $z_{i|j}$ and $z^T_{i|j}$ lie in the range [0,1]. Kullback- Leibler (KL) divergence is then applied to minimize the conditional probability distributions.

\begin{equation}
     \mathcal{L}_{div}=\sum_i \sum_{j\ne i} z^T_{j|i} \text{log}\left( \frac{z^T_{j|i}}{z_{j|i}}\right). 
\label{loss_pgt}
\end{equation}

The KD loss is $\mathcal{L}_{\textit{KD}}=\mathcal{L}_{p-gt}+\mathcal{L}_{div}$. The total loss is combined with GAN loss for the unsupervised learning.

\begin{equation}
     \mathcal{L}_{unsuper}=\mathcal{L}_{f}+\mathcal{L}_{r}+\mathcal{L}_{\textit{KD}}= \mathcal{L}_{f}+\mathcal{L}_{r}+\mathcal{L}_{p-gt}+\mathcal{L}_{div}. 
\label{loss_pgt}
\end{equation}
\begin{figure}[bt!]
\begin{center}
\includegraphics[width=1.0\linewidth]{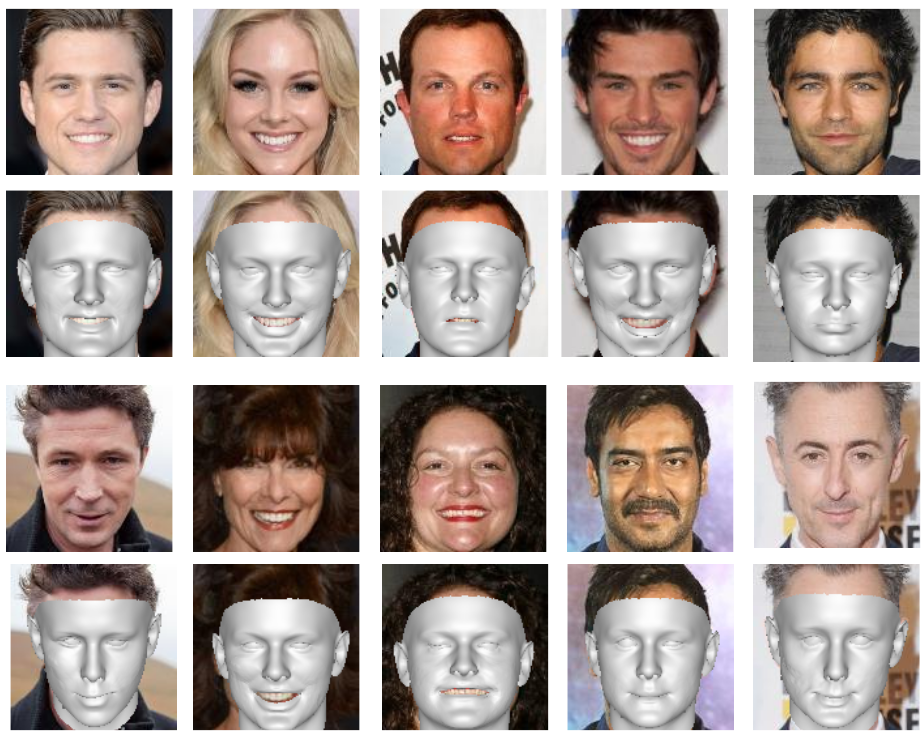}
\end{center}
\vspace{-10pt}
  \caption{\textbf{Samples of 3D fitted 3D face in Voxceleb-3D.} We overlay the fitted 3D face with associated 2D images to show the fitness. Voxceleb-3D contains wider to thinner face shape variation.}
  \vspace{-7pt}
\label{voxceleb-3d}
\end{figure} 

\section{Experiments and Results}
\label{sec:exper}
\textbf{Datasets} 
We use the Voxceleb-3D dataset described in Sec. \ref{sec:methods/supervised}.
There are about 150K utterances and 140K frontal face images from 1225 subjects. 
The train/test split for Voxceleb-3D is the same as \cite{NEURIPS2019_eb9fc349}:
Names starting with A-E are for testing, and the others are for training. 
We manually pick the best-fit 3D face models for each identity as reference models for evaluations. We display samples of 3D fitted face model in Fig. \ref{voxceleb-3d}.

\textbf{Data Processing and Training.} We follow \cite{NEURIPS2019_eb9fc349} to extract 64-dimensional log mel-spectrograms with window size 25 ms, and perform normalization by mean and variance of each frequency bin for the utterance. Input audio clips are randomly crop to 3 to 8 seconds for training. The face images are also normalized by pixel value mean and variance. The face image size of groundtruth and the generator output is 64$\times$64. We also follow 3D face reconstruction work \cite{guo2020towards} to perform 3DMM parameter normalization for the Voice2Mesh supervised setting. In the unsupervised setting, we bilinearly upsample the size of latent image representation to 120 $\times$120 to fit the input size of pretrained network \cite{guo2020towards}.

Network architectures: $\phi_v$ of supervised and unsupervised settings are the same as voice encoder in \cite{NEURIPS2019_eb9fc349}. For the supervised learning, we adopt linear layers to regress 40-/10- dimension vectors from the voice embedding $v$. For the unsupervised learning, the speech-to-image block utilizes the same structure as GANs in \cite{NEURIPS2019_eb9fc349}, and the image-to-3D-face block follows the teacher expert \cite{guo2020towards} to use MobileNetv2 \cite{sandler2018mobilenetv2}.

For training settings, we implement the Voice2Mesh in PyTorch and use Adam as the optimizer. The learning rate is 2$\times$10$^{-4}$, batch size is 128, and total number of iterations is 50K.  


\textbf{Metrics}
We design several metrics based on lines, points, and regions to evaluate 3D face deformation. Note that we apply pose for aligning images and 3D models in Fig. \ref{voxceleb-3d} to show the fitness. In the following numerical evaluation, we focus on the deformation and do not apply pose for the reference models. 

\begin{figure}[bt!]
\begin{center}
\includegraphics[width=0.4\linewidth]{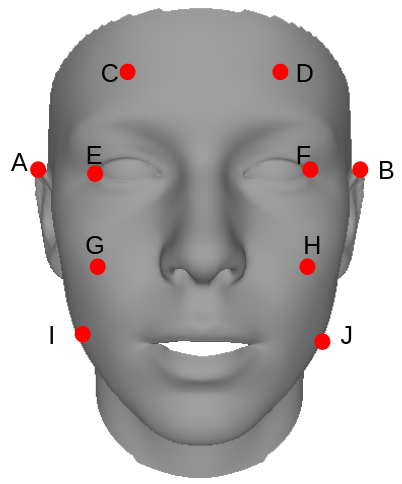}
\end{center}
  \caption{\textbf{Point-to-point distance illustration for our ARE metric.} $\overline{AB}$ is ear-to-ear distance, $\overline{CD}$ is forehead distance, $\overline{EF}$ is outer-interocular distance, $\overline{GH}$ is midline distance, and $\overline{IJ}$ is cheek-to-cheek distance.}
  \vspace{-10pt}
\label{p2pDistance}
\end{figure}

1. Absolute Ratio Error (\textbf{ARE}): Point-to-point distance is usually calculated for measuring 3D face size in aesthetic or surgical applications \cite{sarver2007aesthetic, pallett2010new, abdullah2002inner}. We especially pick representative facial lines as shown in Fig. \ref{p2pDistance} and calculate the distance ratios with outer-interocular distance (OICD). For example, ear ratio (ER) is $\overline{AB}/\overline{EF}$, and the same for forehead ratio (FR), midline ratio (MR), and cheek ratio (CR). These ratios capture the face deformation, such as larger/thinner ER represents wider/thinner faces in terms of eye distances. Then, the absolute ratio error can be calculated as $|\text{ER}-\text{ER}^*|$, where $^*$ denotes the ratio from reference models. This metric is line-based.

2. Normalized Mean Error (\textbf{NME}) of facial landmarks: BFM Face annotates 68 points 3D facial landmark points that lies on the eyes, nose, mouth, and the face outline. We calculate NME of the landmark point set between the predicted and reference 3D face models, i.e. first calculate the Euclidean distance of two landmark sets and then normalize the distance by the face size (square root of face width $\times$ length). This metric is point-based.

3. Point-to-Plane Root Mean Square Error (\textbf{Point-to-plane RMSE}): We perform Iterative Closet Point (ICP) to register predicted and reference 3D face model and then calculate the point-to-plane RMSE. This metric is region-based.

\begin{figure}[bt!]
\begin{center}
\includegraphics[width=1.0\linewidth]{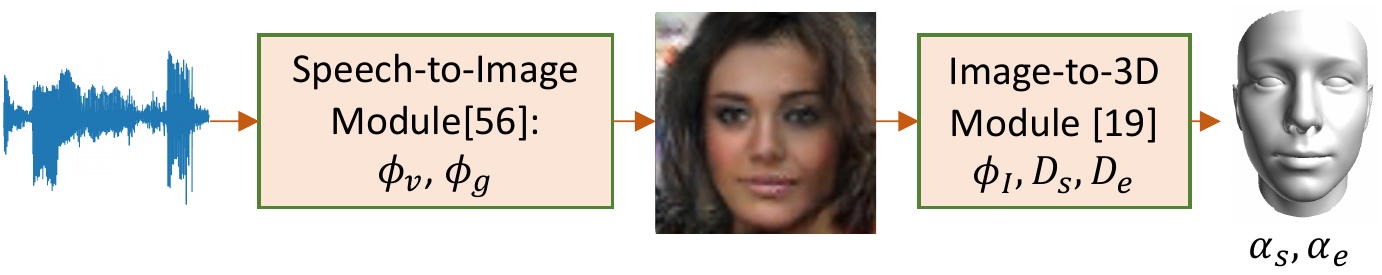}
\end{center}
    \vspace{-5pt}
  \caption{\textbf{Baseline framework.} The baseline is a direct cascade of two pre-trained state-of-the-art modules: one for speech-to-image \cite{NEURIPS2019_eb9fc349} and the other for image-to-3D face generation \cite{guo2020towards}.}
  \vspace{-3pt}
\label{baseline}
\end{figure}

\textbf{Baseline}. 
We build a straightforward baseline by cascading a pre-trained voice-to-image block \cite{NEURIPS2019_eb9fc349} with a pre-trained image-to-3D-face block \cite{guo2020towards} without further joint training.
Both blocks are considered state-of-the-art in the tasks they are responsible of. The illustration is shown in \ref{baseline}.

\subsection{Numerical Results}

We compare numerical results of the baseline (Fig. \ref{baseline}) and our Voice2Mesh under both the supervised and unsupervised learning settings. The results in Table \ref{ARE_metric} exhibit that compared with the baseline of cascade pretrained blocks, either supervised or the unsupervised learning obtains \textbf{15\%-20\% improvements} on the line-based ARE metric. This shows the cross-modal learning is achievable and surpass the performance of direct cascade pretrained blocks remarkably.
The improvement conforms to our intuitions and research from physiology that face shapes are correlated to voices to some extent. Specially, ear ratio (ER) has the largest improvement, which displays that the main contribution of the voice information is getting more accurate wider/thinner head estimation. This also conforms to our instinct that when an unheard speech comes, one can roughly envision whether the speaker face is in overall wider or thinner, but it is comparatively harder for other local facial traits or areas such as the forehead or nose. Through this table, we can quantitatively show how many improvements we can get from joint training with voices.



\begin{table}[tb!]
\begin{center}
  \caption{\textbf{ARE metric study.} Baseline framework is shown in Fig. \ref{baseline}. Our Voice2Mesh for both supervised and unsupervised settings show that voice can help 3D facial geometry prediction and obtain a near 20\% improvement.}
  \vspace{-0pt}
  \label{ARE_metric}
  \begin{tabular}[c]
  {|
  p{1.20cm}<{\centering\arraybackslash}|
  p{1.10cm}<{\centering\arraybackslash}|
  p{1.8cm}<{\centering\arraybackslash}|
  p{2.0cm}<{\centering\arraybackslash}|}
  \hline
      ARE  & Baseline \cite{NEURIPS2019_eb9fc349, guo2020towards}& Voice2Mesh supervised & Voice2Mesh unsupervised  \\
    \hline
       ER & 0.0311 & 0.0157 & 0.0184 \\ 
       FR & 0.0173 & 0.0188 & 0.0172\\ 
       MR& 0.0173 & 0.0172 & 0.0176 \\ 
       CR & 0.0551 & 0.0463 & 0.0484 \\
       \hline
       Mean & 0.0302 & \textbf{0.0245} & \textbf{0.0254}  \\
       & 0\% & \cellcolor{blue!25}\textbf{-18.8}\% & \cellcolor{blue!25}\textbf{-15.9}\% \\
    \hline
  \end{tabular}
  \vspace{-10pt}
\end{center}
\end{table}

\begin{table}[tb!]
\begin{center}
  \caption{\textbf{NME for 3D facial landmark study.} 68 facial landmarks annotations from BFM Face \cite{paysan20093d} is used to select landmark vertices.}
  \vspace{-7pt}
  \label{NME_metric}
  \begin{tabular}[c]
  {|
  p{1.7cm}<{\centering\arraybackslash}|
  p{1.10cm}<{\centering\arraybackslash}|
  p{1.8cm}<{\centering\arraybackslash}|
  p{2.0cm}<{\centering\arraybackslash}|}
  \hline
      Landmark Alignment  & Baseline \cite{NEURIPS2019_eb9fc349, guo2020towards}& Voice2Mesh supervised & Voice2Mesh unsupervised  \\
    \hline
       NME & 0.2969 & 0.2723 & 0.2904 \\
       & 0\% & \cellcolor{blue!25}\textbf{-8.3}\% & \cellcolor{blue!25}\textbf{-2.2}\% \\
    \hline
  \end{tabular}
  \vspace{-14pt}
\end{center}
\end{table}

\begin{table}[tb!]
\begin{center}
  \caption{\textbf{Point-to-Plane RMSE study.} ICP is performed to register the predicted and reference 3D face model. We calculate point-to-plane RMSE after ICP.}
  \vspace{-6pt}
  \label{P2PRMSE_metric}
  \begin{tabular}[c]
  {|
  p{1.7cm}<{\centering\arraybackslash}|
  p{1.10cm}<{\centering\arraybackslash}|
  p{1.8cm}<{\centering\arraybackslash}|
  p{2.0cm}<{\centering\arraybackslash}|}
  \hline
       Model Registration  & Baseline \cite{NEURIPS2019_eb9fc349, guo2020towards}& Voice2Mesh supervised & Voice2Mesh unsupervised  \\
    \hline
       RMSE & 1.348 & 1.210 & 1.312 \\
       & 0\% & \cellcolor{blue!25}\textbf{-10.2}\% & \cellcolor{blue!25}\textbf{-2.7}\% \\
    \hline
  \end{tabular}
  \vspace{-10pt}
\end{center}
\end{table}

\begin{table}[tb!]
\begin{center}
  \caption{\textbf{Part-based point-to-plane RMSE study.} We decompose the holistic face into six parts, including left eye, right eye, nose, mouth, left cheek, and right cheek. We perform ICP for these parts and also calculate the point-to-plane RMSE.}
  \label{part_metric}
  \vspace{-6pt}
  \begin{tabular}[c]
  {|
  p{1.9cm}<{\centering\arraybackslash}|
  p{1.10cm}<{\centering\arraybackslash}|
  p{1.75cm}<{\centering\arraybackslash}|
  p{1.95cm}<{\centering\arraybackslash}|}
  \hline
       Part Registration  & Baseline \cite{NEURIPS2019_eb9fc349, guo2020towards}& Voice2Mesh supervised & Voice2Mesh unsupervised  \\
    \hline
       Left Eye & 0.3945 & \textbf{0.3517} & 0.3779 \\
       Right Eye & 0.3656 & \textbf{0.3349} & 0.3488 \\
       Nose & 0.5250 & \textbf{0.5141} & 0.5177 \\
       Mouth & 0.3435 & \textbf{0.2958} & 0.3149 \\
       Left Cheek & 0.4735 & \textbf{0.4654} & 0.4711 \\
       Right Cheek & 0.5061 & \textbf{0.4916} & 0.4919 \\
    \hline
  \end{tabular}
  \vspace{-10pt}
\end{center}
\end{table}

Then we show NME for 3D facial alignments in Table \ref{NME_metric}. This metric shows smaller gains using voice information since many selected facial landmarks concentrate at eyes, nose, and mouth parts that bear smaller size variations. For example, the nosetip and mid-dorsum usually lies on the center line of faces, and alar base and columella are located around them closely. (See supplementary for details of the terms.)   

Next we show point-to-plane RMSE in Table \ref{P2PRMSE_metric}. This region-based metric reflects the overall registration performance. Supervised Voice2Mesh attains a near \textit{10\%} improvement, and unsupervised Voice2Mesh attains an about \textit{3\%} improvement, compared with the baseline. This evaluation also indicates the capability to reconstruct 3D face models from voices.

Table \ref{P2PRMSE_metric} shows a holistic registration performance, and we also show a part-based breakdown to show point-to-plane RMSE for local regions in Table \ref{part_metric}. Our Voice2Mesh under both supervised and unsupervised settings outperform the baseline in all part registrations. They together contribute to the overall improvement for the holistic registration in Table \ref{P2PRMSE_metric}. 

\textbf{Ablation study}. We study the performance of using different KD methods in the unsupervised learning. We utilize existing KD methods, including vanilla KD \cite{hinton2015distilling}, Attention \cite{zagoruyko2016paying}, SP \cite{tung2019similarity}, Correlation \cite{peng2019correlation}, RKD \cite{park2019relational}, CRD \cite{tian2019crd}, VID \cite{ahn2019variational}, PKT \cite{passalis2020probabilistic}, and train our unsupervised pipeline with different $\mathcal{L}_{\textit{KD}}$. The results are shown in Table \ref{KD_study}. We find that more recent and more advanced KD methods attain similar results, such as RKD, CRD, and PKT have very close performance, compared with earlier methods such vanilla version or using attention map similarity. Therefore, the study validates our conditional probability \cite{passalis2020probabilistic} used in Section \ref{sec:unsuper}. Note that the focus of this work is not on KD. Therefore we conduct extensive survey on the advanced KD methods and exploit their knowledge distillation ability to facilitate the unsupervised voice-to-3D-face learning.   

\begin{table*}[tb!]
\begin{center}
  \caption{\textbf{Study on the different KD methods.} ARE is mused as the metric for comparison.}
  \label{KD_study}
  \begin{tabular}[c]
  {|
  p{1.20cm}<{\centering\arraybackslash}|
  p{1.30cm}<{\centering\arraybackslash}|
  p{1.30cm}<{\centering\arraybackslash}|
  p{1.30cm}<{\centering\arraybackslash}|
  p{1.50cm}<{\centering\arraybackslash}|
  p{1.30cm}<{\centering\arraybackslash}|
  p{1.30cm}<{\centering\arraybackslash}|
  p{1.30cm}<{\centering\arraybackslash}|
  p{1.30cm}<{\centering\arraybackslash}|}
  \hline
      ARE  & Vanilla KD \cite{hinton2015distilling} & Attention \cite{zagoruyko2016paying}& SP \cite{tung2019similarity} & Correlation \cite{peng2019correlation}& RKD \cite{park2019relational} & CRD \cite{tian2019crd} & VID \cite{ahn2019variational} & PKT \cite{passalis2020probabilistic}\\
    \hline
       ER & 0.0306 & 0.0318 & 0.0230 & 0.0227 & 0.0172 & 0.0198 & 0.0213 & 0.0184\\ 
       FR & 0.0173 & 0.0172 & 0.0169 & 0.0173 & 0.0171 & 0.0172 & 0.0172 & 0.0172\\ 
       MR& 0.0173 & 0.0173 & 0.0179 & 0.0179 & 0.0195 & 0.0177 & 0.0178 & 0.0176\\ 
       CR & 0.0540 & 0.0551 & 0.0471 & 0.0471 & 0.0474 & 0.0481 & 0.0471 & 0.0484\\
       \hline
           Mean & 0.0298 & 0.0304 & 0.0262 & 0.0263 & \textbf{0.0254} & 0.0255 & 0.0259 & \textbf{0.0254}\\
    \hline
  \end{tabular}
  \vspace{-17pt}
\end{center}
\end{table*}

\subsection{Visual Results}

\textbf{Compared with VGGFace for real images.}
The advantage of the Voice2Mesh unsupervised learning framework is that both 2D images and 3D face models can be obtained simultaneously. We collect both 2D and 3D faces from Voice2Mesh and then show the generated 2D faces and 3D face models overlapping onto 2D to show the shape fitness. Note that we do not synthesize textures in our Voice2Mesh. This is because we focus on the 3D facial geometry prediction from voices that is physiologically grounded. Texture synthesis from voice is also controversial and does not have a strong support as explained in Sec.\ref{sec:intro}. Therefore, we show a solid gray mesh for visualization. On the other hand, we choose to show unsupervised learning visual results since both 2D and 3D representations are available. Overlay 3D on 2D images helps realize geometry fitness, and 2D textures help recognize identities.  

\begin{figure}[bt!]
\begin{center}
\includegraphics[width=1.0\linewidth]{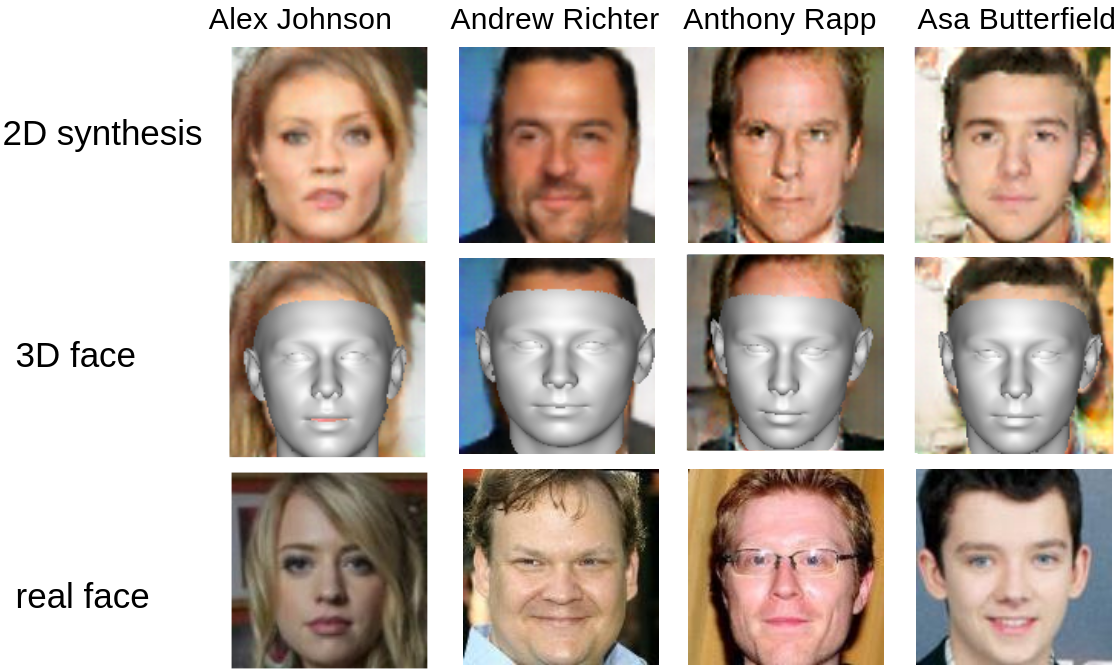}
\end{center}
  \vspace{-10pt}
  \caption{\textbf{Visual Comparison of synthesized faces with VGGFace.} We use unsupervised Voice2Mesh to generate 2D face images and 3D face models. This figure displays that the generated faces are similar to real faces in terms of face shapes. }
  \vspace{-15pt}
\label{VGGFace_compare}
\end{figure}

Real face images from VGGFace are used as reference for identification. The illustration is shown in Fig. \ref{VGGFace_compare}. We can observe that the real face in the second column is wider, and the faces in the third and fourth columns are thinner. The generated 2D face images and 3D face models reflect these shape variations and show similar facial structures. Note that our Voice2Mesh does not predict pose, and we use pose from the pretrained teacher network \cite{guo2020towards} for only visualization purpose to show the fitness. The same usage also appears in later visualization.


\textbf{Prediction Coherence of the same speaker.} We next show the face prediction coherence using different utterances of the same speaker. In Fig. \ref{coherence}, the results exhibit the ability to predict images and 3D face models of similar face shapes for the same speaker. The synthesized 2D face representation contains variations of hairstyles, backgrounds, lights, and textures. However, the 3D face models show consistent geometric prediction to match 2D faces.

\begin{figure}[bt!]
\begin{center}
\includegraphics[width=1.0\linewidth]{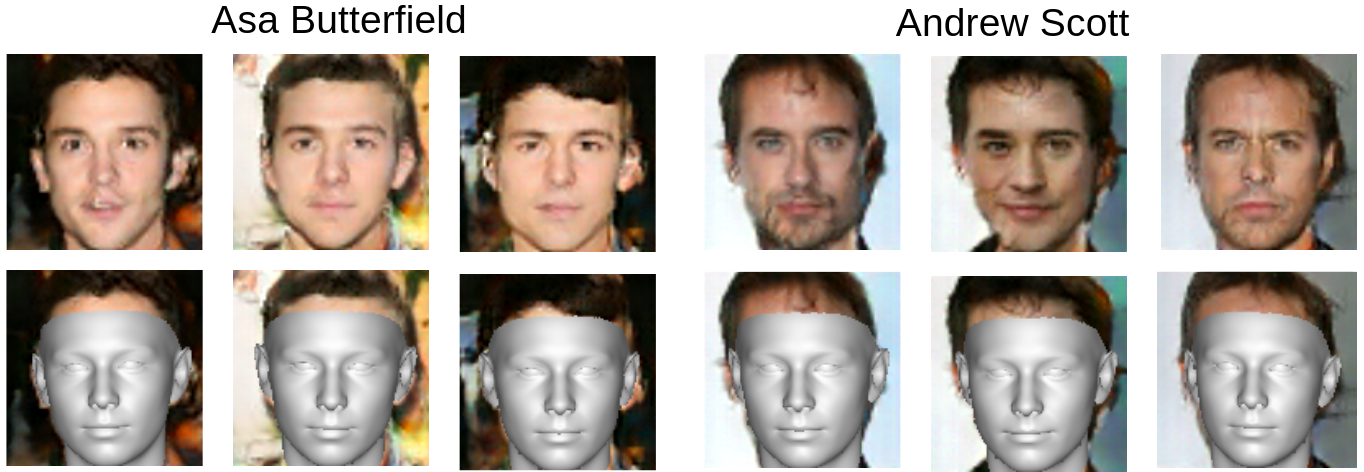}
\end{center}
    \vspace{-10pt}
    \caption{\textbf{Coherence of the inference of 2D and 3D synthesis from the same speaker with different utterances.} }
    \vspace{-7pt}
\label{coherence}
\end{figure} 


\textbf{Visual Comparison with baseline of cascade pretrained blocks.}

\begin{figure}[bt!]
\begin{center}
\includegraphics[width=1.0\linewidth]{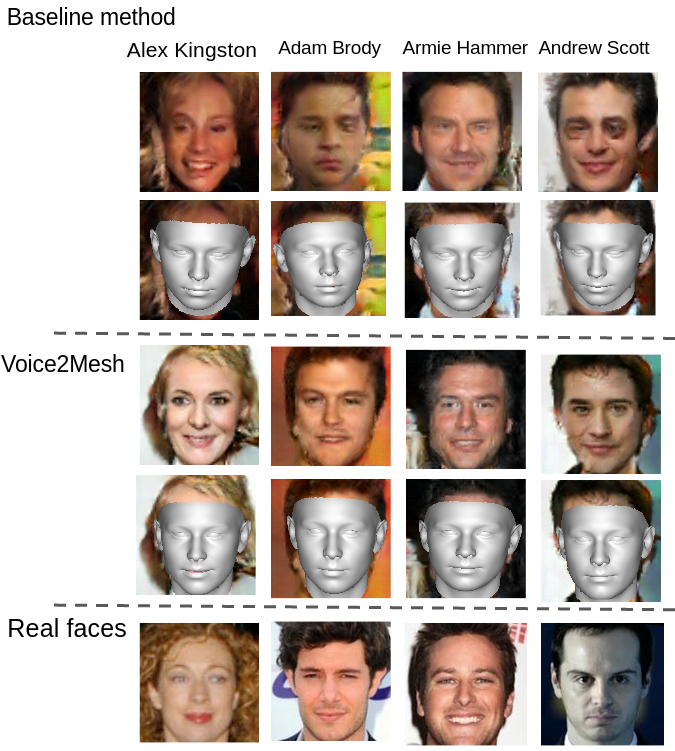}
\end{center}
  \vspace{-10pt}
  \caption{\textbf{Visual comparison of Voice2Mesh and the baseline.} Our Voice2Mesh is able to generate high-quality face images to facilitate the 3D face modeling. We include real face images for reference.}
  \vspace{-7pt}
\label{baseline_comp}
\end{figure}

In Fig. \ref{baseline_comp}, we show comparison of Voice2Mesh with the baseline method (Fig. \ref{baseline}). Images synthesized from the baseline include artifacts, but images synthesized from our Voice2Mesh have higher and more stable quality. We find that joint training of voice-to-image and image-to-3D-face modules in our Voice2Mesh can substantially improve the quality of face image synthesis. Thus, the 3D face models measured in Table \ref{ARE_metric}-\ref{P2PRMSE_metric} have smaller error.



\section{Conclusion and Discussion}
\label{sec:conclusion}
We propose Voice2Mesh to reconstruct 3D face models of speakers given their speech inputs and focuses on the ability of cross-modal learning. We point out that previous 2D face synthesis contains many variations about hairstyles, backgrounds, or facial textures that might draw controversy. Instead of 2D images, we reconstruct 3D face models and focus on the geometry reconstruction based on the supportive studies from physiology. The focus of the work is to analyze whether 3D face models can be predicted from speech inputs only, to quantify the performance gain, and to analyze the source of gain. The goal is not to synthesize high-quality 3D faces that are comparable to those using images or video as the information source.    

We deal with an ideal case when paired voice-to-3D-face datasets exist using supervised learning to learn 3DMM parameters from voice input. Furthermore, we study a real-world scenario when such large-scale paired dataset is absent. We use KD to distill knacks from a teacher expert to a student to facilitate the joint training of voice-to-image and image-to-3D-face networks under this unsupervised learning scenario.

We find that for both settings, voice information can be used to predict 3D face models and attain \textit{15\%}-\textit{20\%} performance gain in terms of line-based metrics, compared with the baseline of cascade pretrained blocks. 

We find that the source of performance gain is from ear-to-ear distance prediction, which corresponds to our observations that we can imagine approximate shapes that whether the speaker face is wider or thinner. 

Voice2Mesh also has improvements based on the other point-based or region-based metrics. From the part registration breakdown, we can see that all part registrations are more accurate and together contribute the holistic face registration improvement.

Besides, we find that our unsupervised learning can also lead to better quality of synthesized 2D face images due to the exploitation of both 2D and 3D representations.

Overall, in this work we try to answer an interesting question, is it possible to reconstruct 3D faces from voices? We present a detailed analysis to show that 3D faces can be roughly modeled from voices to further advance the development of cross-modal learning.




{\small
\bibliographystyle{ieee_fullname}
\bibliography{main}
}

\clearpage
\newpage
\newpage
\pagebreak
\begin{center}
\textbf{\large Supplemental Materials:}
\end{center}
\setcounter{section}{0}
\setcounter{equation}{0}
\setcounter{figure}{0}
\setcounter{table}{0}
\setcounter{page}{1}
\makeatletter
\renewcommand{\theequation}{S\arabic{equation}}
\renewcommand{\thefigure}{S\arabic{figure}}
\renewcommand{\thetable}{S\arabic{figure}}
\renewcommand\thesection{\Alph{section}}
\renewcommand\thesubsection{\thesection.\Alph{subsection}}


\begin{figure}[hbt]
\begin{center}
\includegraphics[width=1.0\linewidth]{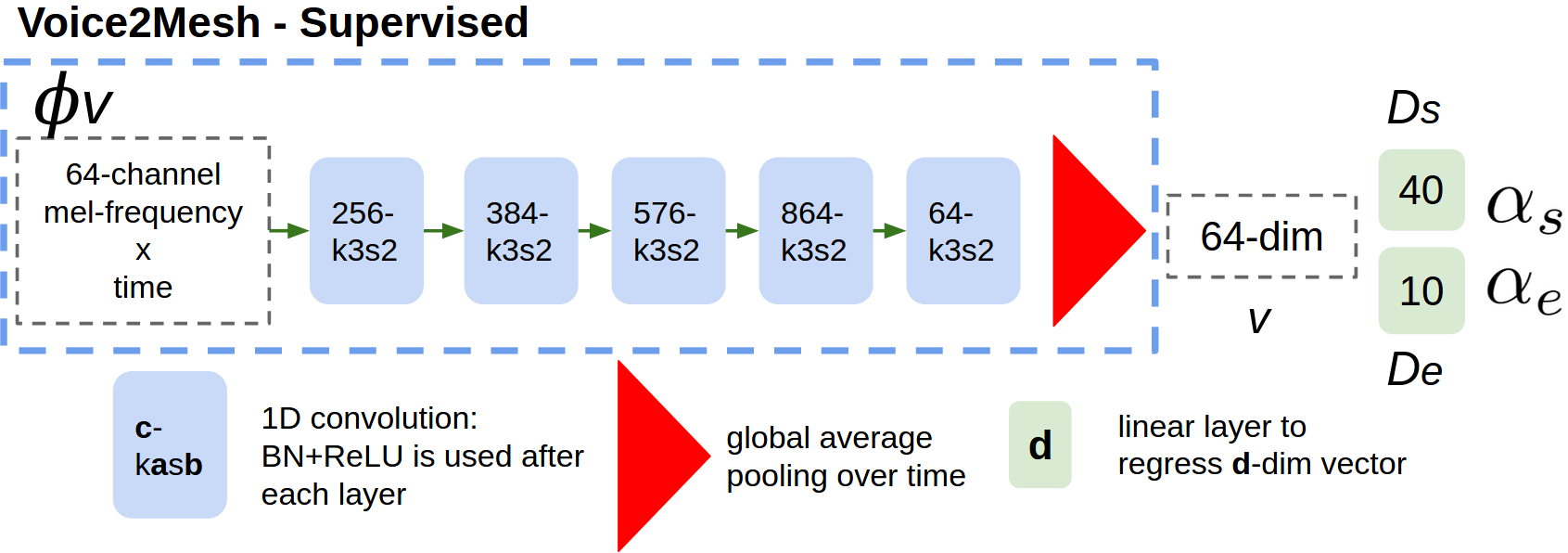}
\vspace{3pt}
\includegraphics[width=1.0\linewidth]{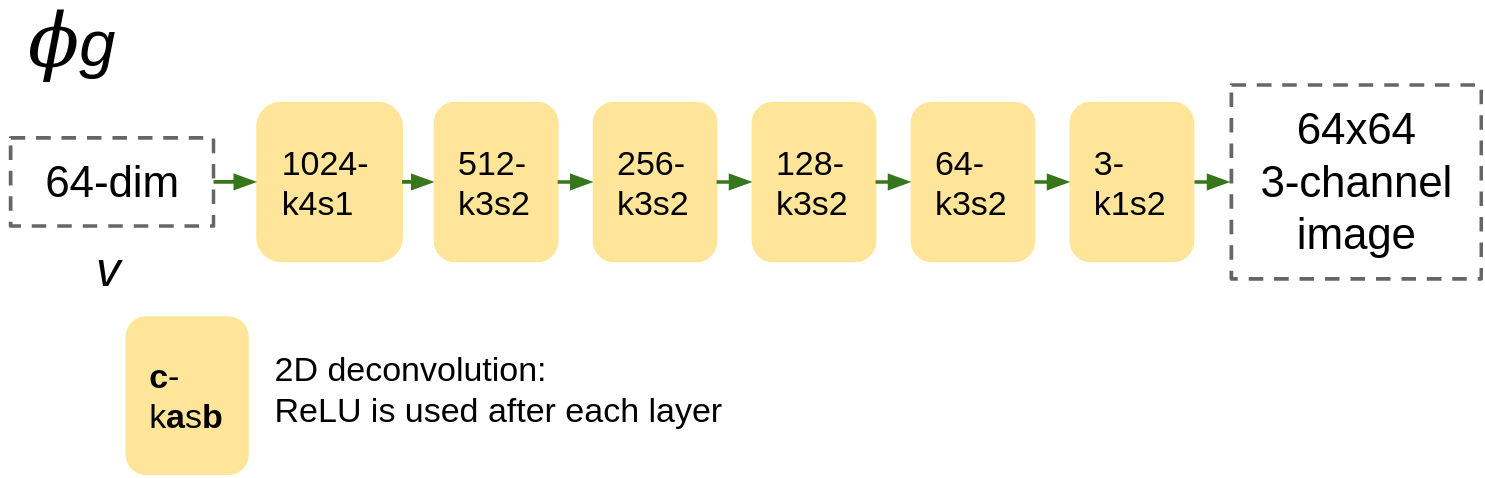}
\vspace{3pt}
\includegraphics[width=1.0\linewidth]{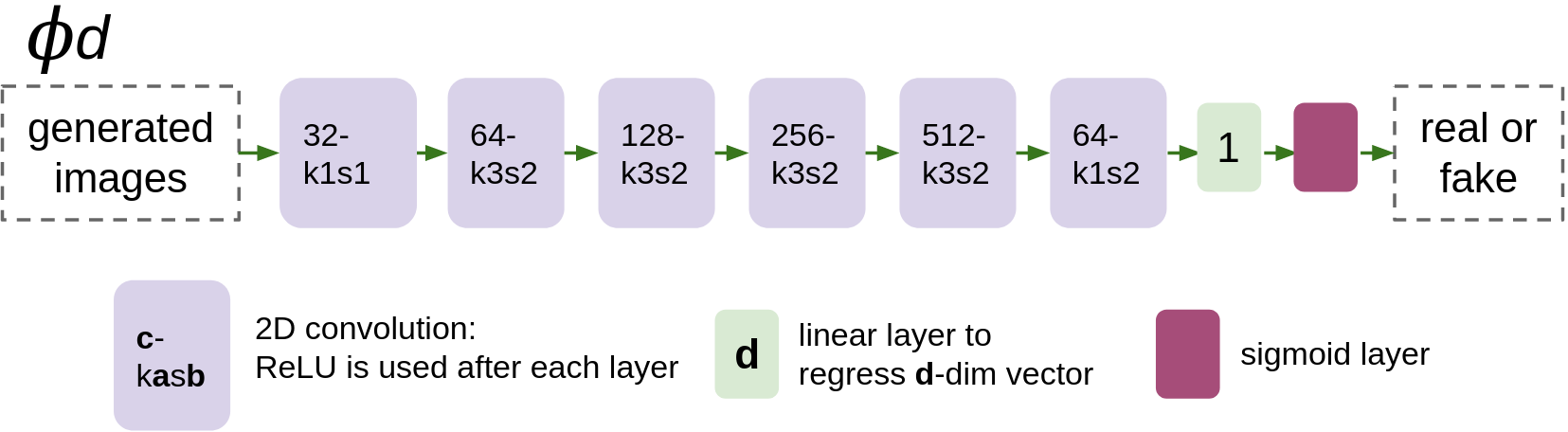}
\vspace{3pt}
\includegraphics[width=1.0\linewidth]{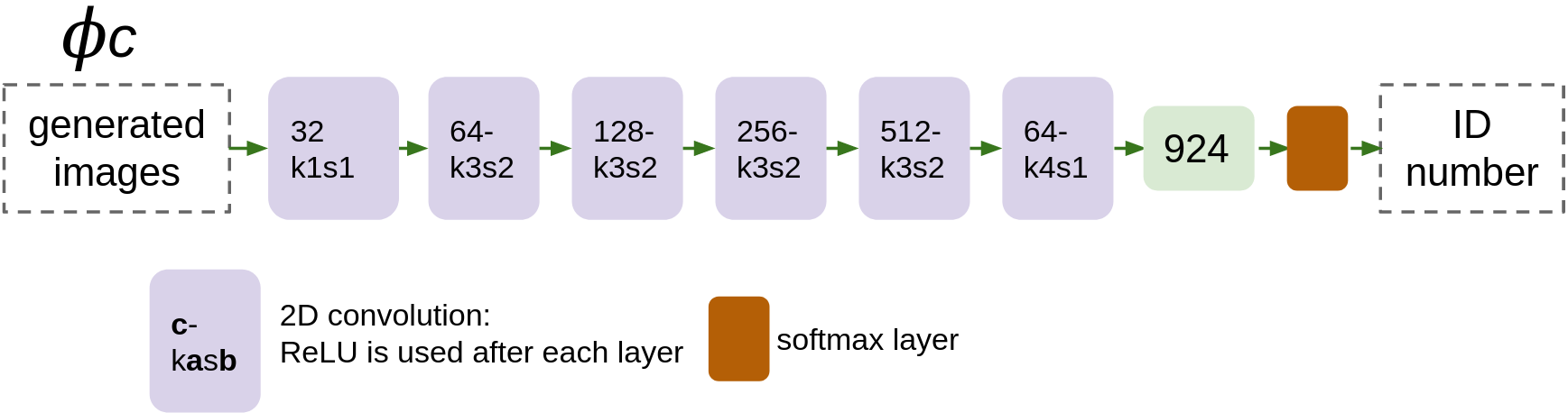}
\vspace{3pt}
\includegraphics[width=1.0\linewidth]{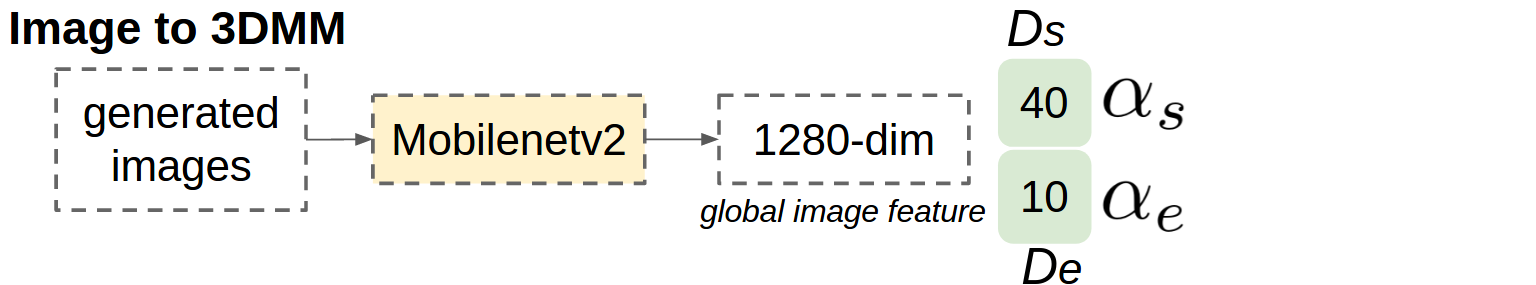}
\end{center}
  \vspace{-6pt}
  \caption{\textbf{Illustrations of network architectures.} \textbf{c}-k\textbf{a}s\textbf{b} means a convolutional layer with a \textbf{c}-channel output using a kernel size \textbf{a} and stride \textbf{b}. \textbf{d} in the linear layer means to output a \textbf{d}-dimension vector. The first graphic contains our Voice2Mesh supervised learning framework with $\phi_v$, $D_s$, $D_e$. The unsupervised setting contains $\phi_g$, $\phi_d$, $\phi_c$, $\phi_v$ (the same structure as in the supervised setting), and the image-to-3DMM module. }
\label{network_arch}
\end{figure} 

\section{Overview}

This supplementary document is organized as follows. 
In Sec.\ref{sec:net_arch}, we show detailed network architectures of our Voice2Mesh under both supervised and unsupervised scenarios. In Sec.\ref{sec:dataset}, we describe details about our Voxceleb-3D training and evaluation split, which complements the contents in Sec.3.2 and Sec.4-Datasets of the main paper. In Sec.\ref{sec:facialPoints}, we explain terms in the main paper describing particular facial points. In Sec.\ref{sec:vis_supervised}, we visualize our generated mesh results from the supervised learning scenario of Voice2Mesh. In Sec.\ref{sec:study_unsupervised}, we present more studies on the unsupervised setting of our Voice2Mesh. 
In Sec.\ref{sec:human_judgement}, we further conduct a subjective preference test to quantitatively compare artifacts generated from our Voice2Mesh and the baseline (cascade pretrained blocks of \cite{NEURIPS2019_eb9fc349, guo2020towards}).
In Sec.\ref{sec:applications}, we especially describe more on the applications of the cross-modal learning of the voice-to-3D-face task.
\vspace{-3pt}

\section{Network Architecture}
\label{sec:net_arch}
We exhibit detailed network architectures used in our Voice2Mesh in Fig. \ref{network_arch}.

\begin{figure}[b]
\begin{center}
\includegraphics[width=1.0\linewidth]{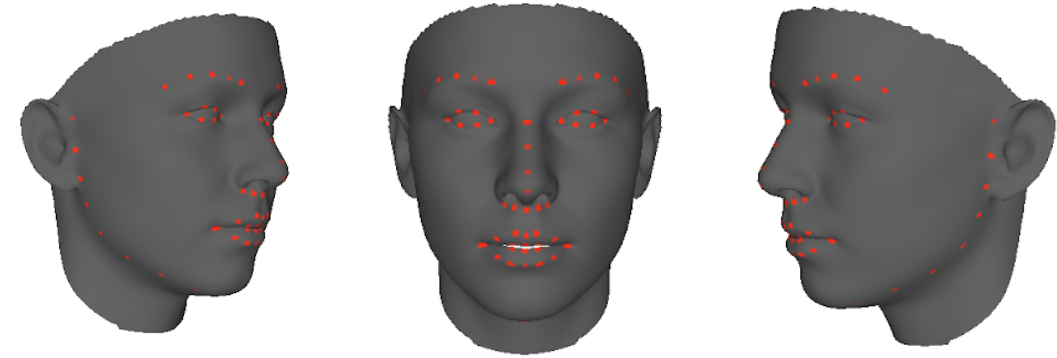}
\end{center}
  \vspace{-14pt}
  \caption{\textbf{Illustration of 68-point facial keypoints.} Here we draw locations of 68-point facial keypoints in BFM Face. These keypoints are used for evaluation in the main paper Sec.4 to compute normalized mean error (NME). }
\label{landmark}
\end{figure} 

\begin{figure}[bt!]
\begin{center}
\includegraphics[width=0.3\linewidth]{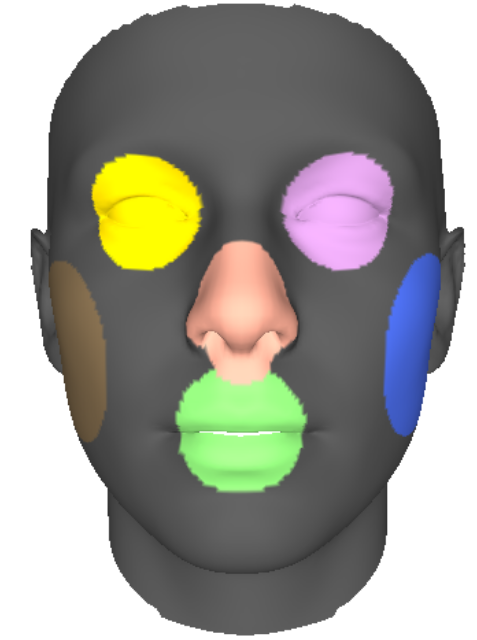}
\end{center}
  \vspace{-14pt}
  \caption{\textbf{Illustration of regions.} Here we show regions of left eye, right eye, nose, mouth, left cheek, and right cheek that are used in the part-based ICP evaluation in the main paper Table 4. }
\label{regions}
\vspace{-5pt}
\end{figure} 

\begin{figure}[bt!]
\begin{center}
\includegraphics[width=0.55\linewidth]{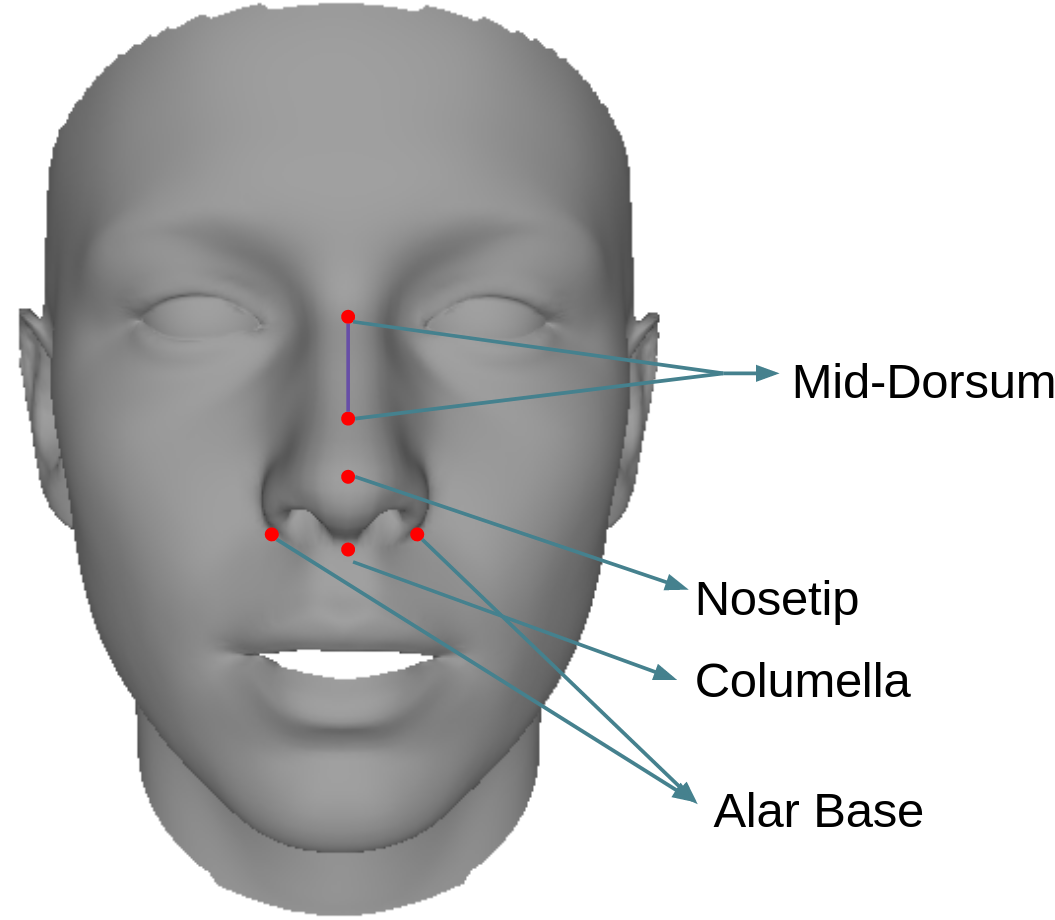}
\end{center}
  \vspace{-15pt}
  \caption{\textbf{Illustration of terms in Sec.4.1 of the paper}. }
\label{terms}
\vspace{-1pt}
\end{figure} 

\section{Voxceleb-3D Training/Evaluation Split}
\label{sec:dataset}
\begin{table}[tb!]
\begin{center}
  \caption{\textbf{Voxceleb-3D training/evaluation split.} Images are not used for numerical evaluation, and thus we mark its number as '-'. We also illustrate gender pie charts below the table.}
  \label{statistics}
  \begin{tabular}[c]
  {|
  p{3.5cm}<{\centering\arraybackslash}|
  p{1.4cm}<{\centering\arraybackslash}|
  p{1.4cm}<{\centering\arraybackslash}|}
  \hline
       Dataset  & Training & Evaluation  \\
    \hline
       \# of utterances & 113K & 0.9K  \\
       \# of face images & 107K & -  \\
       \# of 3DMM param & 107K & 301  \\
       \# of male/female & 485/439 & 182/119 \\
       \# of identity & 924 & 301 \\
    \hline
  \end{tabular}
  \vspace{-20pt}
\includegraphics[width=1.0\linewidth]{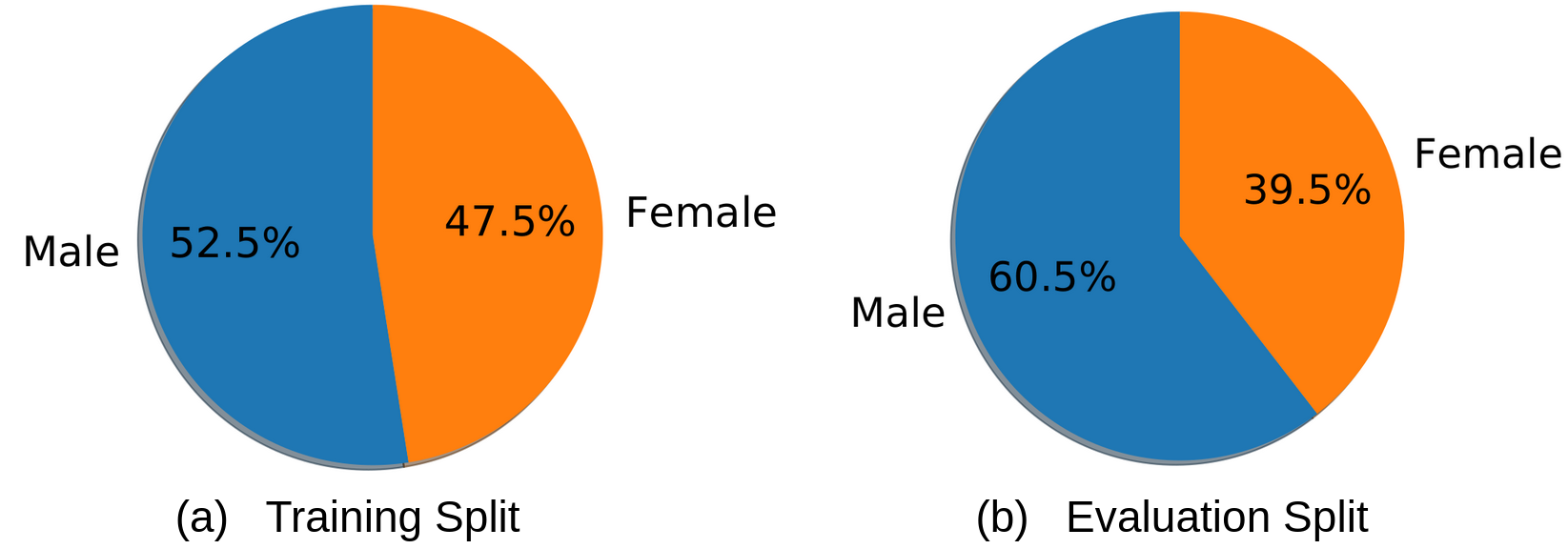}
\end{center}
\end{table}

We display the details of the training/evaluation set split in Table \ref{statistics}. The evaluation set contains 3 utterances per identity, amounting to 0.9K utterances in total.
Face images are not included in the evaluation set because they cannot be used to calculate the reconstruction error of a 3D face model, and thus we put a '-' mark in the table.
For 3D faces, we fit landmarks from images and obtain the optimized 3DMM and reconstructed 3D faces, as described in Sec.3.2 and Sec.4 of the main paper. There are several images for an identity in VGGFace; thus, many 3D faces associated with an identity are reconstructed from images. To create \textit{reference 3D face models} for each identity to conduct numerical evaluation, we manually select one neutral 3D face from the pool that best fit face shape on images. Therefore, there are 301 neutral 3D faces for each identity as the reference. 

During the test time, three utterances for each identity are used as inputs to reconstruct 3D faces. Those three predicted models are then involved to compute numerical performance with the picked one reference model for each identity.

\section{Illustrations of Facial Points and Regions}
\label{sec:facialPoints}
First, we visualize in Fig. \ref{landmark} the locations of the 68-point facial landmarks used to compute NME in the paper. 
Then, we illustrate in Fig. \ref{regions} the areas of facial parts used in paper Sec.4.1 for local ICP registration. 
Lastly, we show in Fig. \ref{terms} the definition of some physiology terms of faces that are used in paper Sec.4.1.

\section{Visualization of Supervised Voice2Mesh}
\label{sec:vis_supervised}

\begin{figure*}[bt!]
\begin{center}
\includegraphics[width=0.7\linewidth]{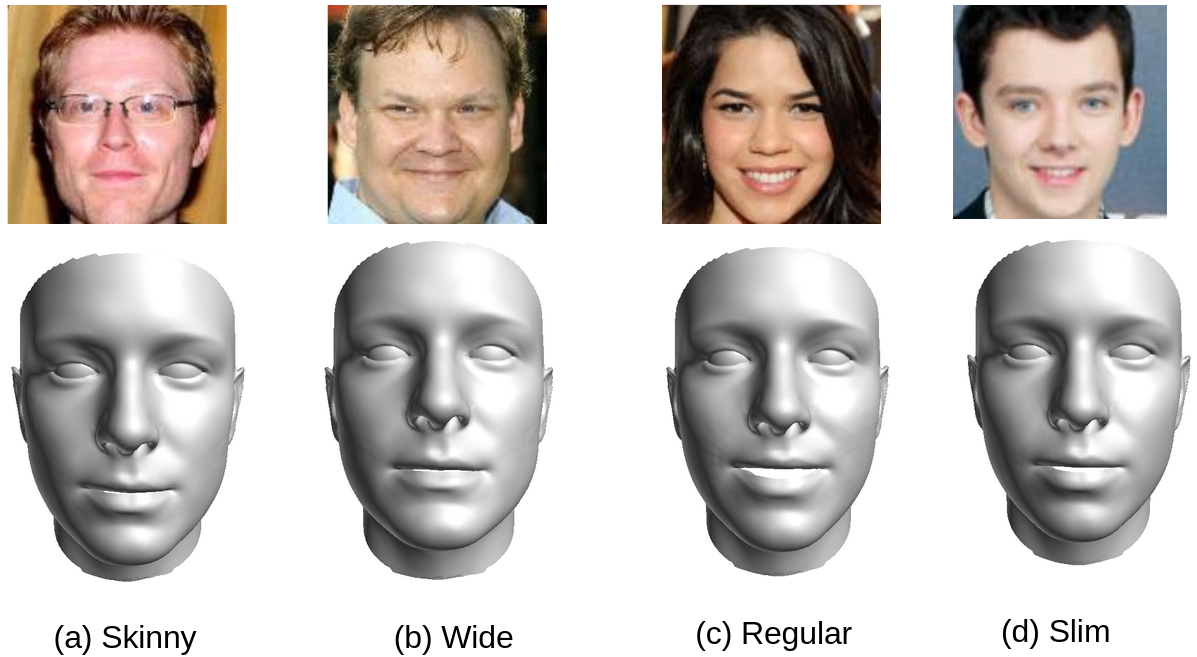}
\end{center}
  \vspace{-6pt}
  \caption{\textbf{Visualization of 3D face models trained by the Voice2Mesh supervised learning setting}. We display four face shapes, skinny, wide, regular, and slim, and their reference images to show the shape correspondence. }
\label{supervised_gt}
\end{figure*}

\textbf{Comparison with a reference face.} In Fig. \ref{supervised_gt}, we present four cases of face shapes -- skinny, wide, regular, and slim -- and show the reference images. The generated face meshes from our supervised Voice2Mesh exhibit the model's ability to generate the corresponding face shapes. This analysis also validates the findings in Table 1 of the main paper: that the lowest absolute ratio error is ear-to-ear ratio (ER) distance, which is associated with overall face shapes indicating wider or thinner faces.

\textbf{Coherence of the same speaker at test time.} We also research inference coherence of face meshes from the same speaker using our supervised Voice2Mesh in Fig. \ref{coherence}. We use different utterances from the same speaker as the input, and display the generated meshes. 
The generated meshes are different for the two speakers in Fig. \ref{coherence}, which can be observed from, for example, the jaw widths;
in contrast, meshes generated from different utterances of the same speaker are highly coherent.
These results demonstrate that our training strategy can successfully learn to generate coherent outputs for the same speaker and can generate different topologies for different identities. 
Finally, our experimental results of coherent syntheses also imply advantages over previous methods for synthesizing images from voices \cite{NEURIPS2019_eb9fc349, oh2019speech2face}, whose generation includes variations of background, hairstyles, etc. Our coherent syntheses exclude these variations and focus on the geometry to validate the correlation between face shapes and voices. 

\textbf{Comparison against the baseline.} We further compare against the baseline that cascades the pretrained blocks, as illustrated in Fig. 6 of the paper. One reference face image, one 3D face mesh generated by our Voice2Mesh, and one by the baseline are presented in Fig. \ref{supervised_comp}.
For its left example, the person of interest has wider jawbone; mesh generated from our Voice2Mesh also shows a similar trait. On the right side, the image shows a wider face shape and apparent cheeks. Our 3D model also displays a similar shape, but the baseline method shows a much thinner face. 

In summary, we use the above visual results to show that our supervised learning for Voice2Mesh is effective. As shown in the numerical results in Table 1 of the main paper, ER distance is the lowest and shows the highest improvements. In Fig. \ref{supervised_gt}, the generated face meshes have similar overall face shapes to the reference images, which shows the model's ability to attain the lowest ER. In Fig. \ref{supervised_comp} the generated face models are more similar to the reference than the baseline in terms of overall face shapes, which validates the ER improvement.

\begin{figure*}[bt!]
\begin{center}
\includegraphics[width=1.0\linewidth]{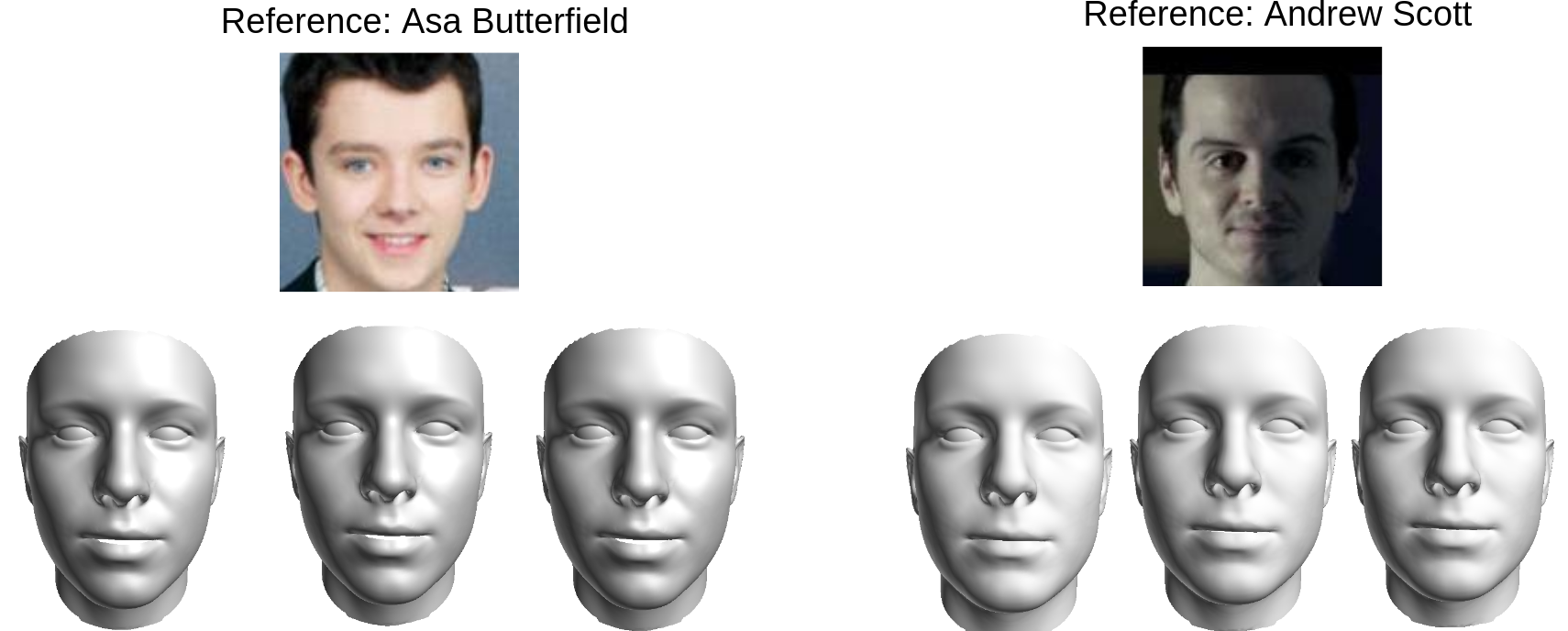}
\end{center}
  \vspace{-6pt}
  \caption{\textbf{Inference coherence of meshes from our supervised Voice2Mesh.} }
\label{coherence}
\end{figure*} 

\begin{figure*}[bt!]
\begin{center}
\includegraphics[width=0.7\linewidth]{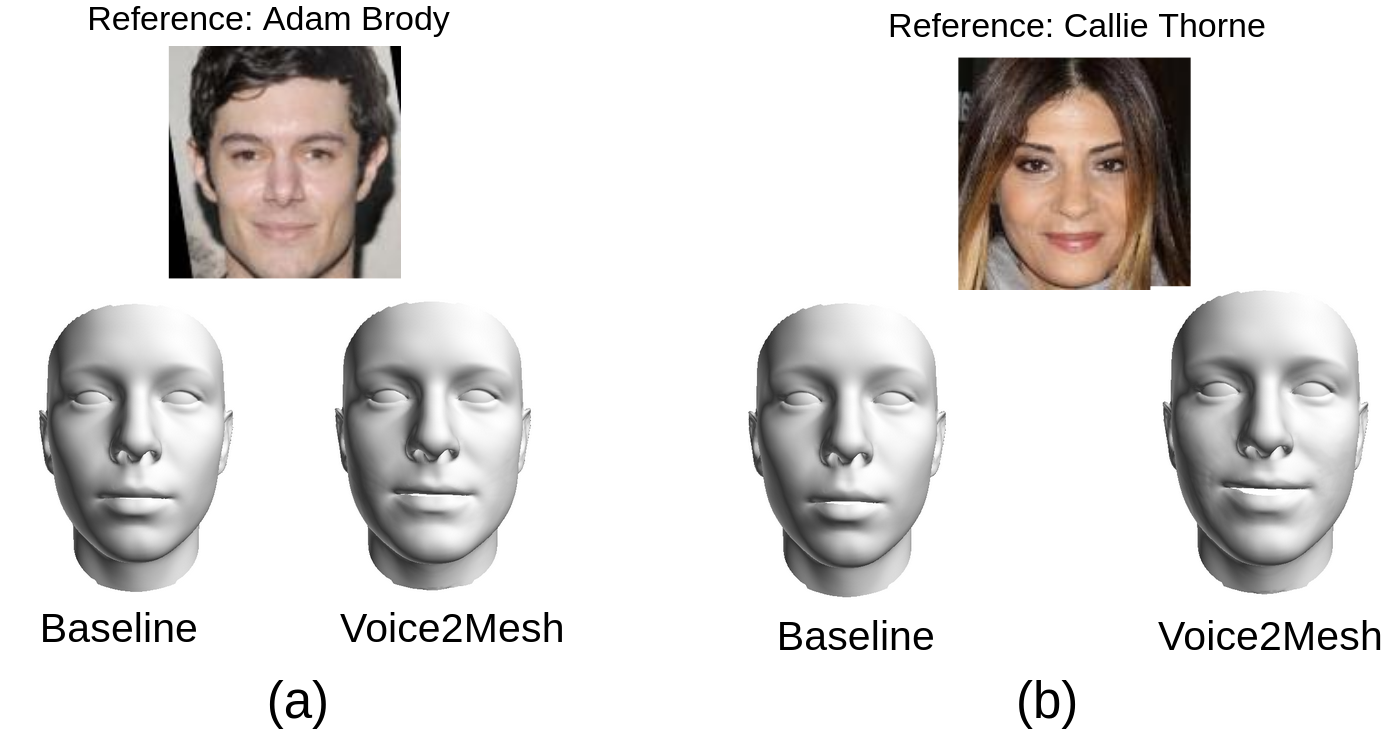}
\end{center}
  \vspace{-6pt}
  \caption{\textbf{Comparison between our generated 3D face models from the supervised setting of Voice2Mesh and the baseline}. Our model in case (a) shows a more squared face with wider jawbone, but the baseline method only shows a slim face. Reference face in (b) is wider and bears apparent cheeks, and our result is much more similar to the reference. }
\label{supervised_comp}
\end{figure*} 

\section{More studies of Unsupervised Voice2Mesh}
\label{sec:study_unsupervised}
Our unsupervised pipeline can generate both 2D images and 3D face models. Although we prefer 3D face models to 2D images in terms of the ability to reconstruct facial geometry, the unsupervised Voice2Mesh relies on the generated 2D images for creating better 3D face meshes. 
We have displayed the better image synthesis quality from our unsupervised Voice2Mesh in Fig. 9 of the main paper. We further follow previous works for face image generation from voices \cite{NEURIPS2019_eb9fc349, oh2019speech2face} to quantify the image synthesis quality. We adopt a well-pretrained gender classification network for images on a much larger dataset \cite{Rothe-IJCV-2018}, and calculate the classification accuracy. Using images generated from the baseline attains 94.86\% accuracy. On the other hand, using images from our Voice2Mesh attains a higher accuracy of 95.32\% . The higher classification accuracy indicates better image synthesis quality using our unsupervised Voice2Mesh. This is because our Voice2Mesh exploits 3D information from the teacher network to secure better face image synthesis.

Further, we illustrate more visual results and comparisons in Fig. \ref{unsupervised_comp_supp} in addition to Fig. 9 in the main paper.

\newcommand {\Hnull} {\mathcal{H}_0}
\newcommand {\Halt} {\mathcal{H}_1}

\section{Subjective Evaluations}
\label{sec:human_judgement}
We further conducted subjective preference tests over the generated outputs to quantify the difference of preference.
The test was divided into three sections, considering \textit{images}, \textit{3D models}, and \textit{joint materials}. Note that we included image generated from our unsupervised learning setting for the test. Though we favor 3D face models more than 2D images to focus on reconstructibility of geometry, better-quality image synthesis also leads to higher-performing 3D mesh generation (as mentioned in Sec.\ref{sec:study_unsupervised}). Therefore we also conducted preference tests for face shape on images. 

\textbf{Evaluation design.} Thirty pairs were included in the test, and 154 subjects with no prior knowledge of our work were invited to the test.
In the first section, each of the ten questions consist of three images-- a reference face image, a face image generated from our method, and a face image generated from the baseline method.
The order of the generated images is randomized.
The subjects were asked to select the face image "whose shape is geometrically more similar to the reference face".
In the second section (10 questions), a similar design was laid out, but 3D face models generated from baseline and our method were used instead of images.
Finally, in the third section (10 questions), each of the two options comprised of a face image and a 3D face model; the subject was asked to consider jointly over the two materials about whose shape better fits the reference.

\begin{figure}[t!]
\begin{center}
\includegraphics[width=1.0\linewidth]{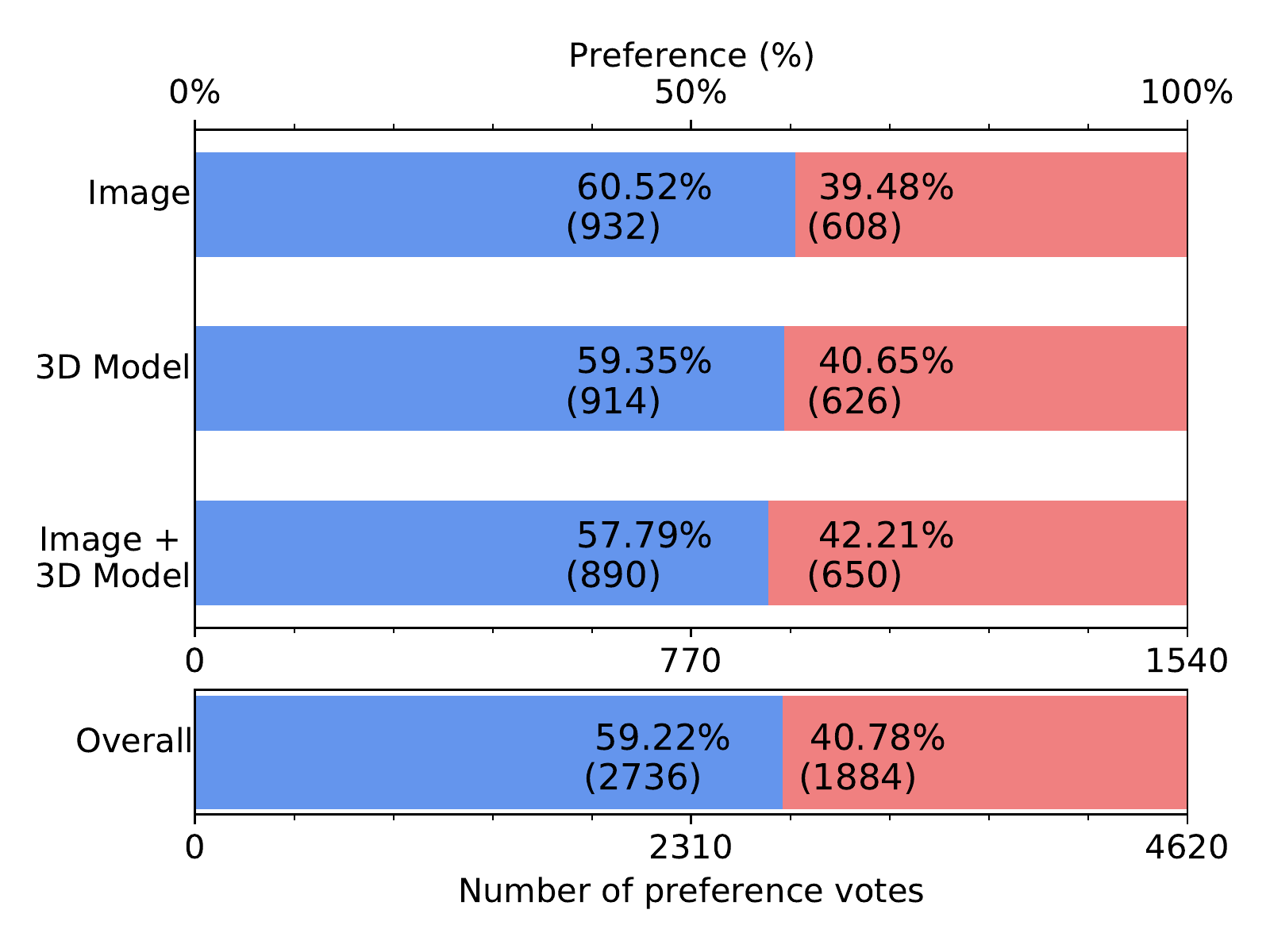}
\begin{tabular}[c]
  {|
  p{1.05cm}<{\centering\arraybackslash}|
  p{1.0cm}<{\centering\arraybackslash}|
  p{1.45cm}<{\centering\arraybackslash}|
  p{1.45cm}<{\centering\arraybackslash}|
  p{1.1cm}<{\centering\arraybackslash}|}
  \hline
         & Image & 3D Model & Image + 3D Model & Overall  \\
    \hline
       $p$-value& \begin{small}$\sim$10$^{-16}$ \end{small} & \begin{small}$\sim$10$^{-14}$ \end{small} & \begin{small}$\sim$10$^{-10}$ \end{small} & \begin{small}$\sim$10$^{-16}$ \end{small} \\
    \hline
  \end{tabular}
\end{center}
  \vspace{-6pt}
  \caption{\textbf{Result of subjective preference tests.} 
  The blue bars are the preference for our method while the red bars are the preference for the baseline method.
  The percentages are labeled on the bar and the total number of votes are enclosed in the parentheses.
  The x-axis on the bottom labels the total number of responses, and that on the top denotes percentage. The $p$-values of the statistical significance tests are provided under the bar. $\sim$ shows the value's order of magnitude.}
\label{fig:pref-test}
\end{figure}

\begin{figure}[bt!]
\begin{center}
\includegraphics[width=1.0\linewidth]{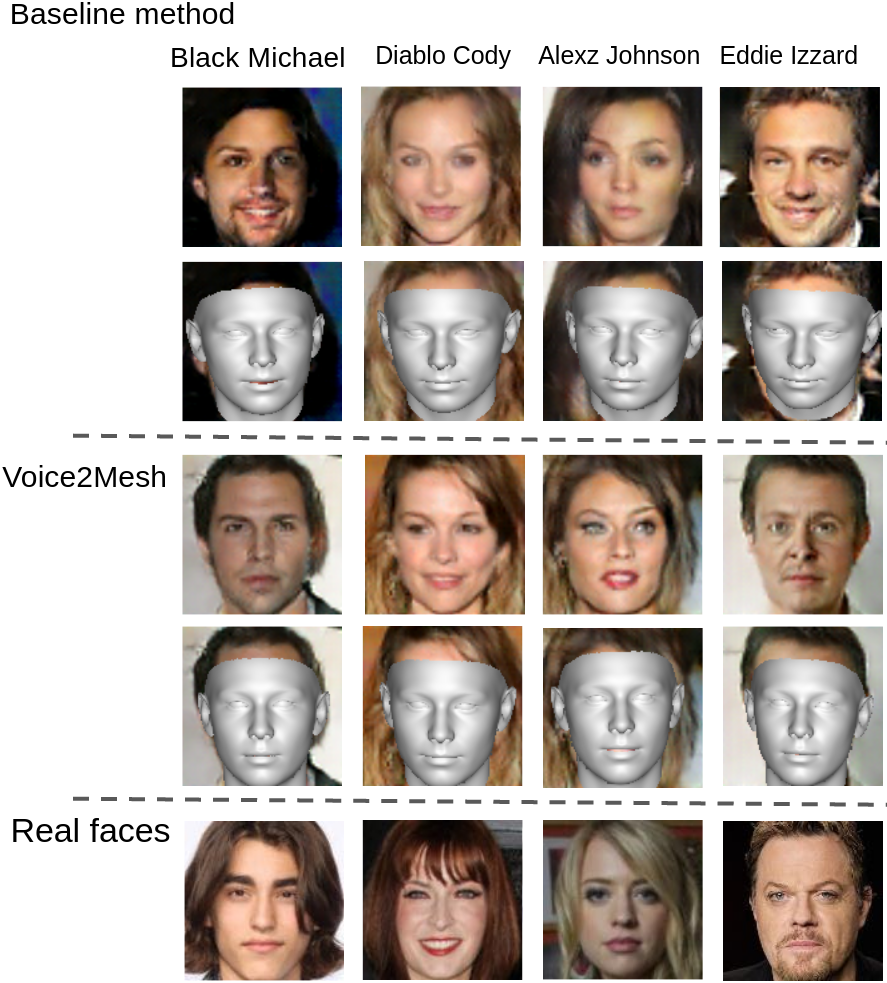}
\end{center}
  \vspace{-6pt}
  \caption{\textbf{Visual comparison of Voice2Mesh from the unsupervised learning setting with the baseline}. This figure presents more results in addition to Fig. 9 in the main paper. }
\label{unsupervised_comp_supp}
\end{figure}

\textbf{Statistical significance test.} Fig. \ref{fig:pref-test} summarizes our subjective evaluation.
We conducted statistical significance test with the following formulation.
The response of a question from a subject is considered as a Bernoulli random variable with parameter $p$.
The null hypothesis ($\Hnull$) assumes $p \le 0.5$, meaning that the subjects do not prefer our model.
The alternative hypothesis $\Halt$ assumes $p > 0.5$, meaning that the subjects prefer our model.
For each section, there are 154 subjects and 10 responses per subject.
For a significance level $\gamma=0.001$, let $b_{n,p}(\gamma)$ denote the quantile of order $\gamma$ for the binomial distribution with parameters $p$ and $n$.
We can decide whether the subjects prefer our model by
\begin{equation}
\begin{split}
    &\mathrm{Reject ~\Hnull ~versus~ \Halt \Leftrightarrow} ~np \ge b_{n,p}(1 - \gamma) \\
    & \Hnull: p \le 0.5, \Halt: p > 0.5.
\end{split}
\end{equation}

As shown in Fig. \ref{fig:pref-test}, $np$ is well above the threshold $b_{n=1540,p=0.5}(1 - \gamma) = 831$, rejecting $\Hnull$ and suggesting that the subjects significantly prefer our model over the baseline.
The single-sided $p$-values are displayed under the bar chart. A lower $p$-value means stronger rejection of $\Hnull$. The $p$-values from our tests are much lower than the level 0.001, showing high statistical significance.
In conclusion, the hypothesis test verifies that the subjects indeed favor the outputs from our model.

\section{Applications of Voice-to-3D-Face Task}
\label{sec:applications}
Although our work is an analysis that validates the correlation between voice and 3D facial geometry, there are a variety of applications for which our work has potential. First, our Voice2Mesh can be used for public security, such as reconstructing face shape of the unheard speech of a suspect or a masked robber. Next, it can be leveraged to create a personal avatar in a gaming or virtual reality system; it is helpful to create a rough 3D face model from voice as the initialization and users can refine its shape based on one's preference. Furthermore,
3D face models reconstructed from voices provide another verification method for person identification in addition to speech and image information. 

Overall, our aim is to study whether it is possible to reconstruct 3D faces form voices and to further advance the development of cross-modal learning.

\end{document}